\documentclass[reprint,aip,twocolumn,amsmath,amssymb,showpacs,cha]{revtex4-1}
\usepackage{graphicx}
\usepackage{textcomp}
\usepackage{psfrag}
\usepackage[colorlinks,linkcolor={blue},citecolor={blue},urlcolor={blue}]{hyperref}
\usepackage{dcolumn}
\usepackage{bm}
\usepackage{multirow}
\usepackage{array}
\usepackage{booktabs}
\usepackage{ctable}
\usepackage{upgreek}
\usepackage{epsfig}
\usepackage{mathrsfs}
\usepackage{amssymb}
\usepackage{amsbsy}
\usepackage{color}
\usepackage{cancel}
\usepackage{pifont}
\usepackage{marginnote}

\begin{document}

\title{A unified model for the dynamics of driven ribbon  with strain and magnetic order parameters} 
\author{Ritupan Sarmah} \email{ritupan@mrc.iisc.ernet.in}
\author{ G. Ananthakrishna} \email {garani@mrc.iisc.ernet.in}
\affiliation{Materials Research Centre, Indian Institute of Science, Bangalore 560012, India}

\begin{abstract}
We develop a unified model to explain the dynamics of driven one dimensional ribbon  for materials with strain and magnetic order parameters. We show that  the model equations  in their most general form explain several results on driven magnetostrictive metallic glass ribbons such as the period doubling route to chaos as a function of a dc magnetic field in the presence of a sinusoidal  field, the quasiperiodic route to chaos as a function of the sinusoidal field for a fixed dc field,  and induced and suppressed chaos in the presence of an additional low amplitude near resonant sinusoidal field. We also investigate the influence of a low amplitude near resonant field on the period doubling route. The model equations also exhibit symmetry restoring crisis with an exponent close to unity. The model can be adopted to explain certain results on magnetoelastic beam and martensitic ribbon under sinusoidal driving conditions. In the latter case, we find interesting dynamics of a periodic one orbit switching between two equivalent wells as a function of an ac magnetic field that eventually makes a direct transition to chaos under resonant driving condition. The model is also applicable to magnetomartensites and materials with two order parameters. 
\end{abstract}


\maketitle

\begin{quotation}
Modeling dynamics of real physical systems is generally difficult. Such a task can be challenging when a single experimental system exhibits a broad spectrum of dynamical features not usually found in a single system. One such system was experimentally realized almost two decades ago  in studies on driven magnetostrictive metallic glass ribbons. These investigations show that the system exhibits the period doubling  route to chaos as a function of a dc magnetic field for a fixed  sinusoidal  magnetic field  and the quasiperiodic  route to chaos as a function of the sinusoidal magnetic field keeping the  dc field fixed. The authors reported several other dynamical features that includes  suppressed and induced chaos in the presence of an additional small amplitude near resonant sinusoidal field, and  stochastic resonance.  Here we design a model with strain and magnetization as order parameters and show that the model not only captures the first three features but also exhibits rich dynamics. The model equations are sufficiently general that they can be adopted to explain the  dynamical features of driven magnetoelastic beam and driven martensitic ribbon. We also find interesting switching dynamics in the case of  forced martensitic ribbon hitherto not reported.
\end{quotation}

\section{Introduction}

Sinusoidally driven nonlinear oscillators have been used as models for  a host of physical systems such as the charge density wave compounds \cite{Huberman79}, Josephson junctions \cite{Benjacob82, Braiman91}, $p-n$ junctions with a nonlinear element \cite{Jeffries85}, driven nonlinear  mechanical  oscillators \cite{Moon, Moon84} etc.  The rich chaotic dynamics exhibited by driven nonlinear oscillators are also ubiquitous to a wide variety of systems such as turbulence in fluid systems \cite{Gollub75}, chemical oscillators \cite{Swinney85} and   cardiac tissues \cite{Glass83}.
Since chaos is usually an undesirable feature in practical applications, developments in  chaos control\cite{Strogatz, LR}, also realized in driven nonlinear oscillators, have been applied to  control the output of laser system \cite{Roy92}, enhance the performance of  permanent magnet synchronous motor\cite{Liu04} and cardiac arrhythmias \cite{Garfinkel92}. Natural to multistate weakly driven oscillators in the presence of ubiquitous  noise is the stochastic resonance \cite{Benzi81}, another generic noise induced cooperative phenomenon  that enhances the signal to noise ratio. This phenomenon is also realized in several disciplines ranging from  physics (SQUID magnetometers, optical and electronic devises) \cite{Gammaitoni98} to biology \cite{McDonnell09}. 

The fact that variety of physical systems could be represented by driven nonlinear oscillators can be ascribed to universality of the dynamics. However, real physical systems such as the magnetostrictive ribbon are much more complicated. While details of the material properties may not be relevant for reasons of universality,  they  require at  least two order parameters for a proper description. Such systems abound in nature.  For example,   ferroic materials have  co-existence of two or more order parameters such as strain, magnetization, electric polarization etc \cite{Eerenstein06, Kainuma06, Fiebig02}. The application of a field complimentary to one order parameter (OP) induces changes in another OP. For example, ferroelectromagnetic materials are materials that induce  magnetization with the application of an electric field or electric polarization with the application of a  magnetic field. Similarly, the well studied ferroelectric materials exhibit strain and electric polarization. Magnetostrictive materials and ferromagnetic martensites \cite{FM} also belong to the class of materials  exhibiting strain and magnetization. Clearly, a general description of such systems would require at least two order parameters for a proper description. More than one order parameter also suggests that we should expect richer dynamics due to enhanced phase space dimension.

Indeed rich dynamics has been realized in experiments on magnetostrictive metallic glass ribbon  \cite{Vohra91a,Vohra91b, Vohra94,Vohra95,Vohra93,Vohra94a}, almost twenty years ago.   In a series of papers,  Vohra and his group reported a host of dynamical features in the study of strain bifurcations in magnetostrictive metallic glass ribbons subjected to the combined influence of sinusoidal (ac) and dc magnetic fields \cite{Vohra91a,Vohra91b,Vohra93,Vohra94,Vohra94a,Vohra95}. In particular, they reported the quasiperioidic (QP) route to chaos when the amplitude of the sinusoidal applied magnetic field $h_{ac}$   was increased in the presence of an additional dc magnetic field $h_{dc}$,  while they found the period doubling (PD) route to chaos when  $h_{dc}$ was increased keeping the ac field fixed  \cite{Vohra91a,Vohra91b}. They also reported several dynamical features such as the suppression and shift of the period doubling bifurcation point  \cite{Vohra91b}, induced subcritical bifurcation under small amplitude near resonant condition \cite{Vohra94}, suppressed and induced chaos \cite{Vohra95}, and stochastic resonance \cite{Vohra94a}.  These dynamical features are observed in  unannealed metallic  glass samples unlike the  PD and QP routes to chaos reported  in annealed samples earlier\cite{Ditto89}.

Magnetostrictive metallic glasses such as METGLAS alloy 2605S-2($Fe_{78}B_{13}Si_9$) have 
striped domain structure that lies in the plane of the ribbon. The anisotropic nature of magnetostrictive strain and domain movement under applied magnetic field leads to the softening of the elastic modulus that goes by the name $\Delta E$ effect  \cite{Levingston82, Savage86, Squire90}.  The softening is significant even for small applied fields and is also nonlinear, hardening beyond a  critical  value of the magnetic field. The $\Delta E$ effect is  also reversible. The buckling of the magnetostrictive metallic glass ribbons under applied magnetic field (along the length of the ribbon) has been experimentally demonstrated \cite{Savage86}. Most of these investigations  are near equilibrium studies except the one that exploits the nonlinear dependence of the Young's modulus on the magnetic field in the field-annealed samples \cite{Ditto89, Savage90} where the PD and QP routes to chaos have been reported \cite{Ditto89}. However, the authors emphasize that rich dynamics exhibited by the metallic glass samples (listed above) \cite{Vohra91a, Vohra91b, Vohra93, Vohra94, Vohra94a, Vohra95} are not due to the $\Delta E$ effect  \cite{Vohra91a,Vohra91b}  as the samples used are unannealed ribbons where the $\Delta E$ effect is insignificant.  

Some of these features have been explained using appropriate normal forms \cite{Bryant86} by appealing to the  universal nature  of bifurcation phenomenon. However, since this approach  cannot go beyond the first bifurcation, there is a need to develop a model that explicitly includes both elastic and magnetic degrees of freedom. Moreover,  the fact that both the ac and dc fields are  necessary ingredients for most experimental features emphasizes the importance of describing magnetic degrees of freedom along with strain. Part of the difficulty in developing a model in terms of the two relevant OP's can  be traced to the lack of knowledge about the nature of elastic and  magnetic nonlinearities,  and their mutual coupling.   We recently developed a model set of equations for  strain and magnetic order parameters, and preliminary results has been published \cite{Ritupan12}. Here we present detailed investigations of the model to show that it captures  a good number of these experimental features apart from  exhibiting rich dynamics.

The model equations can be adopted to describe the dynamics of a ribbon of a material exhibiting  strain and magnetization under an experimental set-up similar that used in the study of the dynamics of a magnetoelastic beam \cite{Moon, Moon84, Cusumano95}. There are number of variants of the experimental set-up all of which have one end of the beam fixed and the other end is driven. (See Ref. \cite{Moon} for example.)  Here we target the results of an experiment where  both the phase plot and the associated Poincar\'e map has been reported \cite{Cusumano95}. Here a permanent magnet is attached to one end and  the other end is driven by a low frequency sinusoidal magnetic field (less than a Hz)\cite{Cusumano95}. The authors explain the results using a parametric forcing equation for the strain variable.  However, if the ribbon is magnetoelastic, a more natural approach would be to write down the equations of  motion for both the elastic strain and magnetization. As we shall see, our model equations explain the experimental phase plots and the Poincar\'e maps.

Our model can also be adopted to describe driven martensitic ribbon by omitting  the magnetic degrees of freedom.  Such experiments were performed early eighties in the context of internal friction studies of nonmagnetic martensites. The geometry is again similar to the two experiments considered. Experiments have been carried out on nonmagnetic martensite  ($Cu_{82.9}Al_{14.1}Ni_3$) samples in the neighborhood of  martensite start temperature $T_m$ where elastic nonlinearities are strong \cite{Suzuki80, Wuttig81}.   Multiperiodic response has been reported when the amplitude of the imposed vibration exceeds a critical value under near resonant driving conditions. In the absence of magnetic degree of freedom, our equations reduce to the  one dimensional model for martensites introduced by Bales and Gooding \cite{Bales}.   As we shall show,  apart from explaining the multiperiodic oscillations,   we find an unusual dynamics of period one orbit switching between the two equivalent wells that eventually makes a direct transition to chaos.

\section{A Unified  Model} 

Keeping in mind that we need to develop a model that can be adopted to describe the three distinct physical systems, we begin by compiling the  minimal features that must be included in the model.  First,  considering the geometry of the experimental set-up used in the above experiments, we assume a thin long ribbon (or a beam) of length $l$ (larger than the other two dimensions) fixed at one end and  set into oscillations by applying an ac magnetic field at the other end.   Thus, for all practical purposes, one dimensional description would be adequate. Moreover, since the displacement or strain is monitored at the free end, it would be adequate to use  a single strain variable $\epsilon$ (in one dimension) along with the  magnetic order parameter $m$.  Second,  the elastic free energy is  taken to have a sixth order polynomial form in the strain variable since  such a form would  also be consistent with the  normal forms used for explaining certain dynamical features \cite{Vohra91a, Vohra91b, Vohra94, Vohra93, Vohra95}. Moreover, a sixth order polynomial is most appropriate for describing  martensites.  For the case of magnetostrictive metallic glass samples,  such a form can be made consistent  with the speculation by Vohra et al  \cite{Vohra91a, Vohra91b, Vohra94, Vohra93, Vohra95} that weak magnetic nonlinearity  drives the strong elastic nonlinearity by including appropriate magnetoelastic coupling. Third, as for the choice of the  magnetoelastic coupling term,  we  use a weighted sum of symmetry preserving and symmetry breaking terms  of the individual free energies of the two order parameters (for reasons explained below).  Finally,  keeping in mind that we require the general model to be reduceable to the case of martensites, we use a dissipative term that has the form of the Rayleigh dissipation function. Then, as we shall show later, the model reduces to the one dimensional model for martensites introduced by Bales and Gooding \cite{Bales} that supports twinning when the magnetic degree of freedom is ignored. For the case of magnetostrictive ribbon, the dissipation simply serves as a damping mechanism. Finally,  the equation for the strain variable is coupled to an equation for the magnetic order parameter. 

The total free energy of the system has three contributions, namely, the strain free energy, magnetic free energy and magnetoelastic free energy. The total free energy is given by
\begin{equation}
F_T=F_{\epsilon} + F_m + F_{m\epsilon}.
\end{equation}
In one dimension, the strain variable is $\epsilon = \frac{\partial u(y)}{\partial y}$. Then, the elastic free energy is given by 
\begin{equation}
F_{\epsilon}=\int dy'\left(f_{l}(\epsilon(y'))+\frac{D}{2}\big(\frac{\partial \epsilon}{\partial y'}\big)^2 \right),
\end{equation}
where $f_{l}$ is the Landau free energy density and the second term is the gradient free energy with $D$ being a positive constant that can be conveniently set to unity.  Guided by the above considerations,  we use  a sixth order polynomial for the elastic free energy $f_{l} = \frac{\theta}{2} \epsilon^2 -\frac{\beta}{2} \epsilon^4 + \frac{\Delta}{6} \epsilon^6$. 
Here $\beta$ and $\Delta$ are positive constants, but $\theta$ can take on positive or negative values.  The minima of the free energy are located at $\epsilon=0$ and $\epsilon = \pm \epsilon_s$ with $\epsilon_s =[(\beta + \sqrt{\beta^2 -\Delta\theta})/\Delta]^{1/2}$.  The parameters $\beta,\Delta$ (of the order of unity) and $\theta$ are sufficient to control the minima.  When $\theta \ge  3\beta^2/4\Delta$, $\epsilon=0$ is the true minima identified with the high temperature phase. At $\theta =3\beta^2/4\Delta$, we have a first order transition with the free energy vanishing at a transformation strain  $\epsilon^2 =\pm 3\beta/2\Delta$ and $\epsilon=0$.  For $\theta$ negative, $\epsilon=0$ state is unstable. This behavior can be parameterized by using $\theta =  \frac{T- T_s}{T_f - T_s} =\tau_{\epsilon}$ where $T_f$ and $T_s$ are the first and second order transition temperatures respectively.

The magnetic free energy is 
\begin{eqnarray}
\nonumber
F_m &=&\int dy'\big[a\tau_c\frac{m^2}{2} + b\frac{m^4}{4}+\frac{D_m}{2}\big(\frac{\partial m}{\partial y'}\big)^2 \\
&-& m(h_{dc} + h_{ac}sin \omega t) \big].
\end{eqnarray}
Here $m$ is the dimensionless magnetization and $\tau_c = \frac{T- T_c}{T_c}$ with $T_c$ referring to the Curie temperature, the third term is the gradient free energy, $h_{dc}$ and $h_{ac}$ are the dc and ac magnetic fields, and $\omega$ is the frequency of the forcing field.   The constants  $a,b$ and $D_m$ are positive. Here again, the values of the constants $a$ and $b$ can be obtained by using the magnetization at a given temperature. For all practical purposes, the three constants can be set to unity as has been done here.   Now consider  the  magnetoelastic coupling. Generally, the nature of this coupling depends on the change in the crystal symmetry from  the parent to the transformed phase. This approach can not be used here as we  are working in one-dimension. Further, in the case of metallic glass, the nature of the magnetoelastic free energy is not known, particularly since  it has a glassy structure. However, coupling terms either preserve or break the invariance $\epsilon \rightarrow -\epsilon$ and $m\rightarrow -m$ in the elastic and magnetic free energies respectively. In the absence of any information,  we model it  as a weighted sum of symmetry preserving and breaking terms. For the metallic glass sample, the weighted sum is appropriate due to lack of long range order while  the symmetry preserving term is favored for crystalline magnetoelastic materials.  On the basis of this, we take the general form of the magnetoelastic coupling  to be  
\begin{equation}
F_{m\epsilon}=- \frac{\xi}{2}\int dy'[(1-p)\epsilon(y')m(y') + p\epsilon^2(y')m^2(y')],
\end{equation}
where $\xi$ is magnetoelastic coupling coefficient and $ 0 \le p \le 1$ is an adjustable weight factor. Other types of gradient coupling between $\epsilon$ and $m$ are ignored to keep the model simple. 

We introduce a dissipative term in the form of the Rayleigh dissipative function given by 
\begin{equation}
F_{diss}=\frac{\gamma}{2}\int\!\!\left(\frac{\partial \epsilon}{\partial t}\right)^2 dy'.
\end{equation}
The value of the constant $\gamma$ is not easy to estimate and is taken to be a free parameter. While the above form of $F_{diss}$  merely serves as a damping term for the magnetostrictive metallic glass, we emphasize  that this kind of dissipation  is  essential for inducing twinned configuration for martensites\cite{Bales}. Indeed, the dissipative force opposing the abrupt motion of the twin interface is required to establish a mechanical equilibrium \cite{LL}.

The kinetic energy of the system is given by  
\begin{equation}
T = \frac{1}{2} \rho \int \left(\frac{\partial u(y',t)}{\partial t}\right)^2 dy', 
\end{equation}
where $u(y,t)$ is the displacement variable and $\rho$ is the density. 
Then, using the Lagrangian $L= T - F_T$ and the Lagrange equations of motion 
\begin{equation}
\frac{d}{dt}\left(\frac{\delta L}{\delta \dot{u}(y)}\right)-\frac{\delta L}{\delta u(y)} =  {-\frac{\delta F_{diss}}{\delta \dot{u}(y)}},
\end{equation}
we get
\begin{eqnarray}
\frac{\partial^2 \epsilon(y)}{\partial t^2} &=& \frac{\partial^2}{\partial y^2} \Big[ \tau_{\epsilon} \epsilon(y)-2 \beta \epsilon^3(y)+ \Delta \epsilon^5(y)- \frac{\partial^2}{\partial y^2} \epsilon(y) \nonumber \\
 &+&\gamma\frac{\partial \epsilon(y)}{\partial t} -\xi \Big(p \epsilon(y) m^2(y)+ \frac{(1-p)}{2} m(y)\Big)\Big]. 
\label{S}
\end{eqnarray}

For the magnetic order parameter we use dissipative dynamics   given by 
\begin{equation}
\frac{\partial m}{\partial t} = - \Gamma \frac{\delta F_T}{\delta m}.
\end{equation}
Using this and the total free energy $F_T$, we get
\begin{eqnarray}
\label{M}
\frac{\partial m(y)}{\partial t}&=&-\Gamma \Big[\tau_c m(y) + m^3(y) -  \frac {\partial^2 m(y)}{\partial y^2}- h_{ac} \sin \omega t  \nonumber \\
&-& h_{dc} - \xi\Big( p \epsilon^2(y) m(y)+ \frac{(1-p)}{2}\epsilon(y)\Big)\Big],
\end{eqnarray}
where $\Gamma$ is a suitable time constant. Note that the above equation does not represent a pure relaxational dynamics since $m(y)$ is subject to ac and dc drives.  Indeed, $\epsilon(y)$ is being driven through $m(y)$ that is itself subject to a sinusoidal field. We note that the free energies used for  magnetization and magnetoelastic coupling are such that they can represent electric polarization as well.

We shall now simplify these equations to suit the experimental set-up where one end is fixed and the other end is driven.  We note that since the amplitude  is monitored at the free end, the fundamental mode that is excited by the applied magnetic field is such that the length of the ribbon $l$ is equal to half the wavelength of the fundamental mode. We assume that only the dominant mode of vibration is supported. This is equivalent to using  $\epsilon(x,t)=A(t)sin \,\pi x/2l$ and $m(x,t)=B(t)sin \,\pi x/2 l$ in Eqs.(\ref{S}, \ref{M}).  
Transforming  Eqs.(\ref{S}, \ref{M}) into rescaled variables ($k=\pi/2l, \tau=kt, \Omega= \omega/k, \Gamma'=\Gamma/k$) and using only the dominant mode we get 
\begin{eqnarray}
\nonumber
\ddot{A}(\tau)&=&-\Big[\tau_{\epsilon} A(\tau)+3 \beta A^3(\tau)-\frac{5\Delta}{8}A^5(\tau)+k^2A(\tau) \\
\label{A}
&&+\gamma k\dot{A}(\tau)+\frac{\xi}{2} \Big( 3 pA(\tau)B^2(\tau) - (1-p) B \Big) \Big],\\
\nonumber
\dot{B}(\tau)&=&-\Gamma' \Big[\tau_cB(\tau)+\frac{B^3(\tau)}{2}+k^2 B(\tau) -\frac{\xi}{2} \Big( (1-p)A\\
&& +  p A^2(\tau)B(\tau) \Big)- h_{ac} \sin \Omega \tau - h_{dc} \Big].
\label{B}
\end{eqnarray}
The overdot now refers to the redefined time derivative.  Equations (\ref{A}) and (\ref{B}) constitute a coupled set of nonlinear nonautonomous ordinary differential equations with several parameters. We shall use $\tau_{\epsilon}, \tau_c, \gamma,\Gamma', \Omega$ and $\xi$ as free parameters. The parameters $\beta$ and $\Delta$ are chosen appropriately for each case  of the magnetostrictive metallic glass ribbon, the magnetoelastic beam and  the martensite ribbon.  In addition, both $h_{ac}$ and $h_{dc}$ are the experimental drive parameters for the magnetostrictive metallic glass while $h_{ac}$ is a drive parameter for the other two cases. Apart from this, we have no knowledge of the values of the parameters.

\section{Driven magnetostrictive metallic glass ribbon}

We  now use Eqs. (\ref{A}, \ref{B}) to explain several dynamical features of the driven magnetostrictive metallic glass ribbon. In this case, the presence of both the ac and dc magnetic fields appears to be crucial for the observed features. The application of the dc field in the presence of the ac field has two effects. First, the vibrations are centered around the strain amplitude induced by the dc field. Noting that the strain amplitude in Eqs. (\ref{A}, \ref{B}) corresponds to the situation where both ac and dc fields are present, the strain amplitude centered around the dc field can be described by redefining $A$ to be the deviation from the strain amplitude $A_{dc}$ due to the dc field alone, ie., $A \rightarrow A- A_{dc}$. The value of $A_{dc}$ needs to be supplied. Second,  the free energy for the magnetic  order parameter is tilted to one side. Further, the shape of the free energy is also affected due to the presence of nonlinear  magnetoelastic coupling term. In particular,  we note that apart from the  additive $B$ term in Eq. (\ref{A}), $B^2$ appears multiplicatively with $A$. This is equivalent to parametric forcing. This affects the shape of the effective elastic free energy. Thus,   we should expect $\beta$ to be a function of $h_{dc}$.  To understand the nature of the dependence of $\beta$ on $h_{dc}$,   we first note that keeping $\beta =1$ and $h_{dc}=0$, increasing the amplitude of the ac field leads to  a period doubling cascade as expected from studies on Duffing like oscillators  \cite{Musielak05}.  On the other hand, keeping $h_{dc} =0$ and fixing $h_{ac}$ at a value where we find period one, and increasing $\beta$ also leads to a period doubling sequence. In contrast, keeping $h_{ac}$  at a value where chaos is seen (with $\beta=1$) and  increasing $h_{dc}$ reverses the period doubling sequence reflecting the compensating relationship between $\beta$ and $h_{dc}$. This can be made quantitative by  keeping $h_{ac}$ at a suitable value, say  $h_{ac} = 9.2$ corresponding to the onset of periodic four orbit, and increasing both $h_{dc}$ and $\beta$ such that we always retain the onset of period four orbit. This gives a relation $\beta \approx 1+ 10 h_{dc}$. However, while $\beta$ is nearly linear in $h_{dc}$, the prefactor depends on the value of $h_{ac}$ used.   For further calculations, we use this parameterized form of $\beta$ in Eqs.(\ref{A}, \ref{B}).  For simplicity, we further set $\Delta =1$ for this case.

The dynamics of the above system of equations is quite complicated due to the presence of several parameters.  Here we will not attempt to study the dynamics in the multiparameter space which is neither the focus of the present investigation nor desirable in such a large  multiparameter space. However, interesting dynamics can be expected when $\tau_{\epsilon}$ is such that two or three minima exist. We also use $\tau_c < 0$. Thus, we first map-out the region of the parameter space where interesting dynamics is seen.  

\begin{figure}[t]
\vbox{
\includegraphics[height=3.7cm,width=8.0cm]{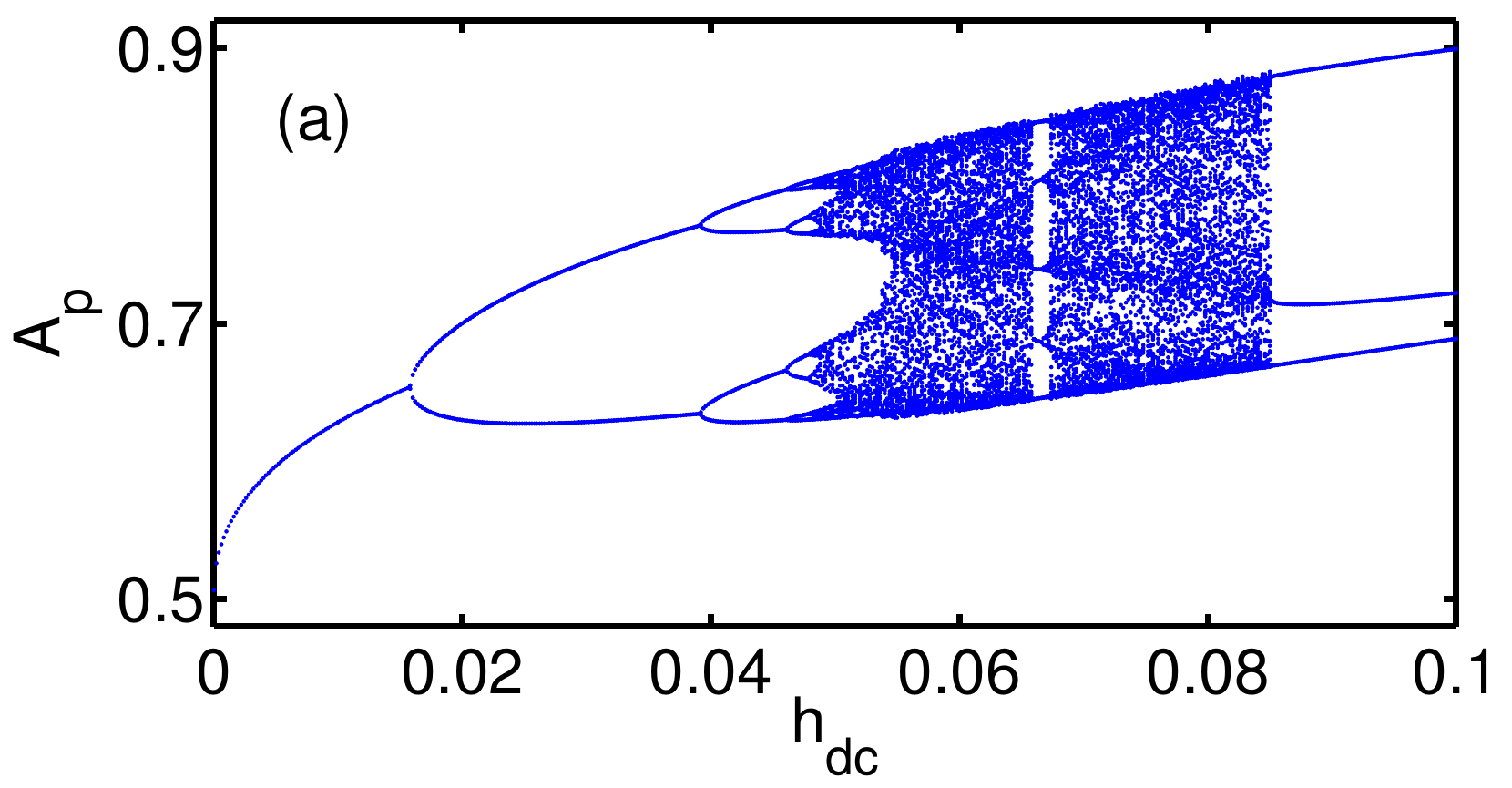}
\includegraphics[height=3.7cm,width=8.0cm]{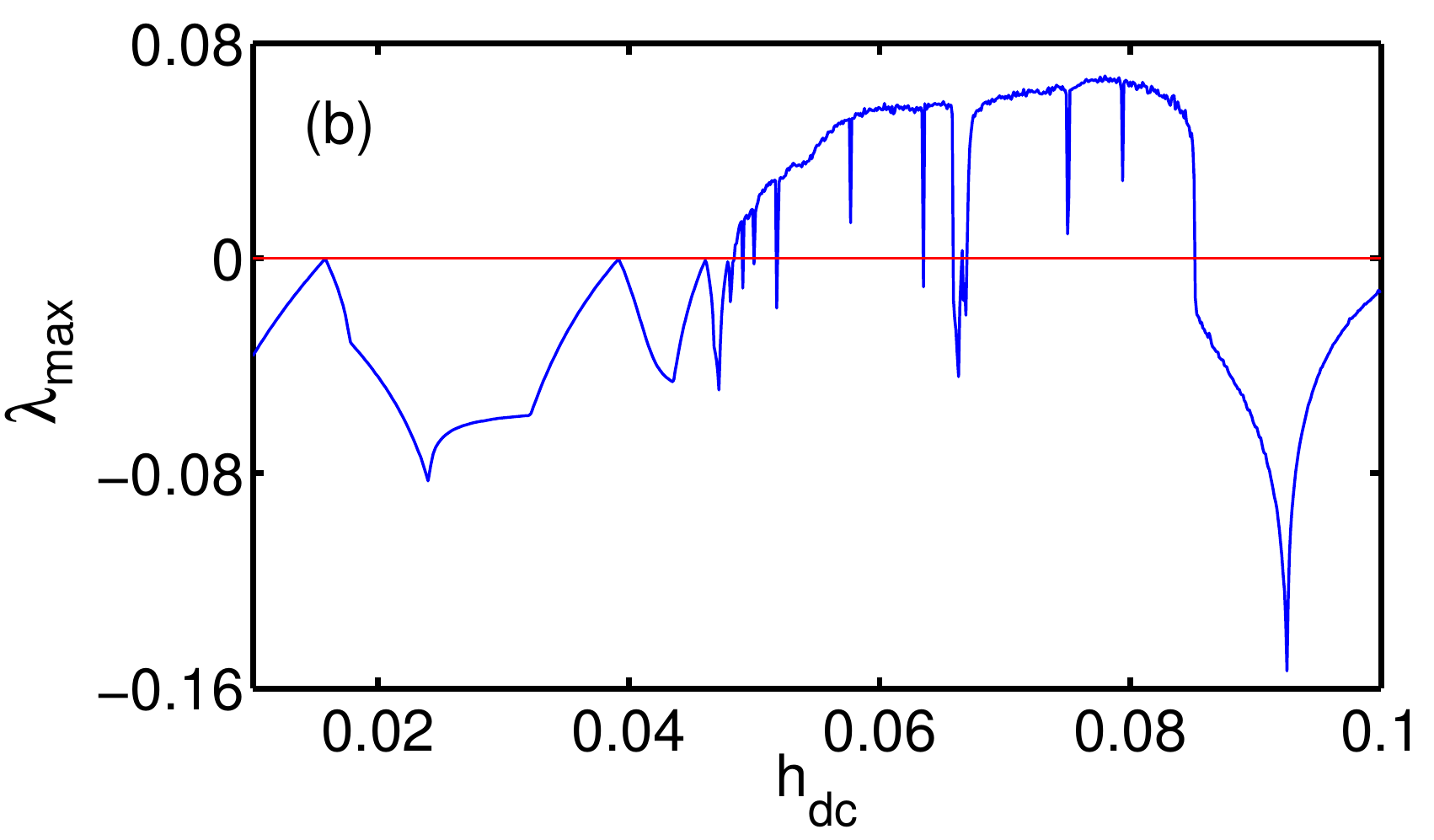}
}
\caption{(a) Period doubling bifurcation as a function of $h_{dc}$ for $\tau_c = -0.2, \tau_{\epsilon}=-1.0,\Omega=1,\xi=0.6, \gamma= 1.592, \Gamma'=0.09, h_{ac}=10.5$, and $p=0.32$. (b) The corresponding largest Lyapunov exponent for the same set of parameters. }
\label{PD-Bif}
\end{figure}

We begin by investigating the  possibility of realizing the period doubling route to chaos when $h_{dc}$ is increased for a fixed  $h_{ac}$. Here, we keep $\tau_{\epsilon}=-1$ so that $\epsilon=0$ state is unstable although interesting dynamics is also observed for $\tau_{\epsilon} < 3\beta^2/4$.  Other parameters are fixed at $\tau_c = -0.2,\Omega = 1,\gamma= 1.592, \Gamma'= 0.09, l =5, \xi=0.6$ and $p=0.32$.  All quantitative measures of the dynamics are obtained after discarding the first $1.5 \times 10^5$ time steps.  Then, increasing $h_{dc}$ keeping $h_{ac} = 10.5$  leads to the period doubling  route to chaos. Bifurcation diagrams have been constructed by sampling the strain amplitude $A$ stroboscopically. A plot of the period doubling sequence is shown in Fig. \ref{PD-Bif}(a). The associated largest Lyapunov exponent is shown in Fig. \ref{PD-Bif}(b). The PD route is seen for a range of  parameter values around the values used for Fig. \ref{PD-Bif}.

\begin{figure}
\vbox{
\includegraphics[height=4.0cm,width=7.5cm]{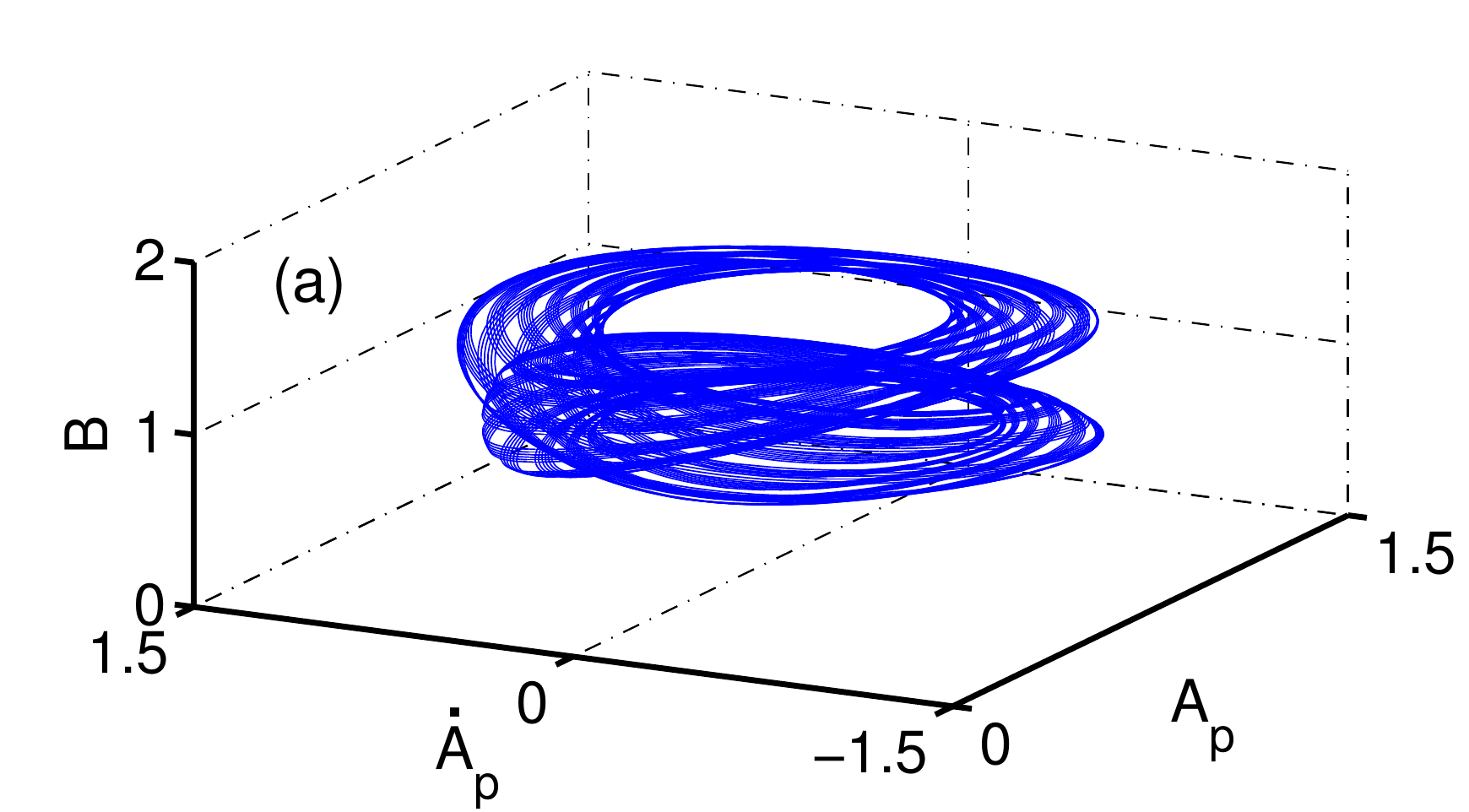} 
\includegraphics[height=3.2cm,width=7.5cm]{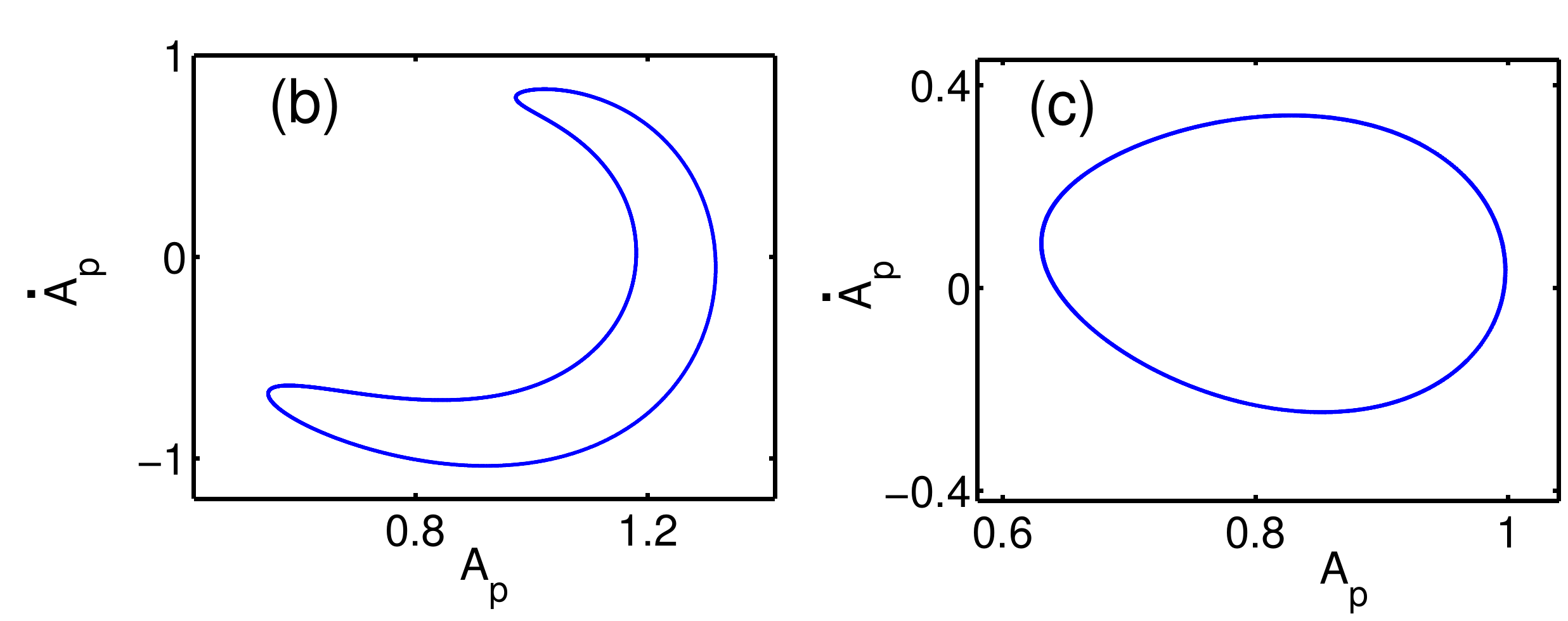}
\includegraphics[height=3.5cm,width=7.5cm]{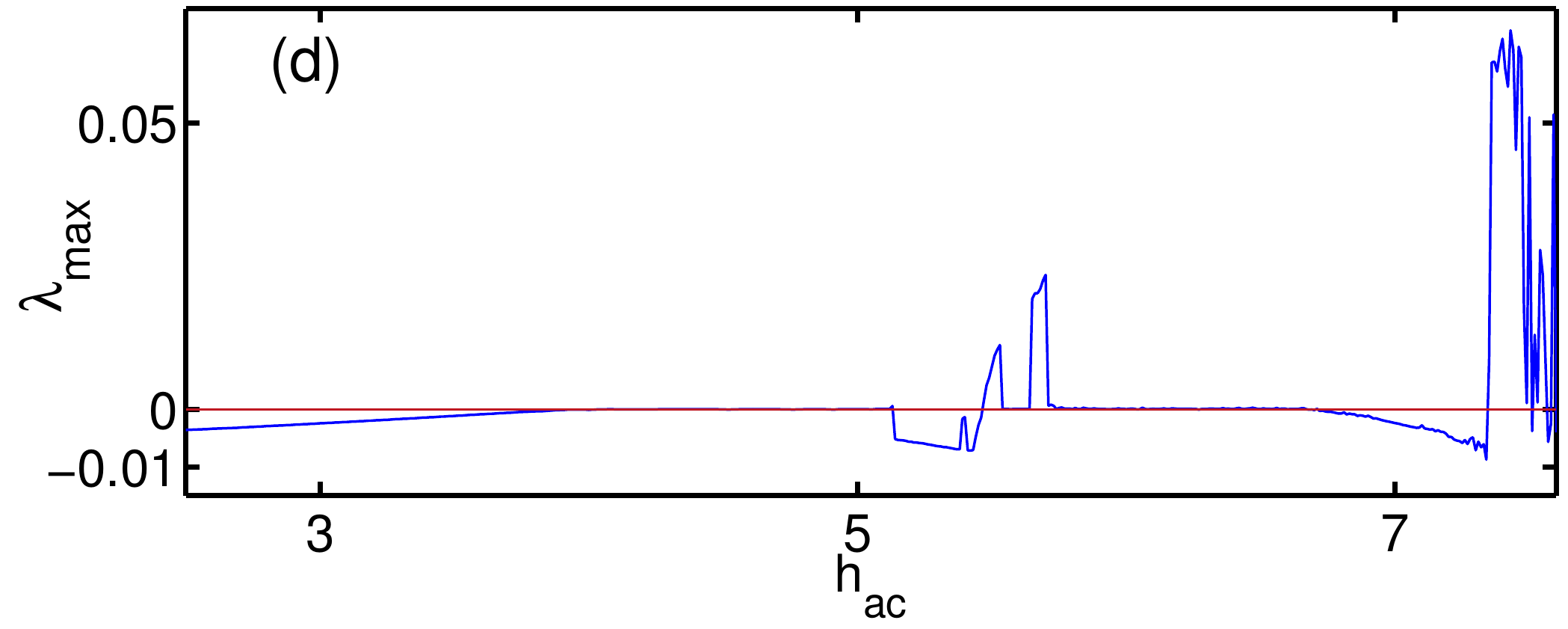} 
\includegraphics[height=3.2cm,width=7.5cm]{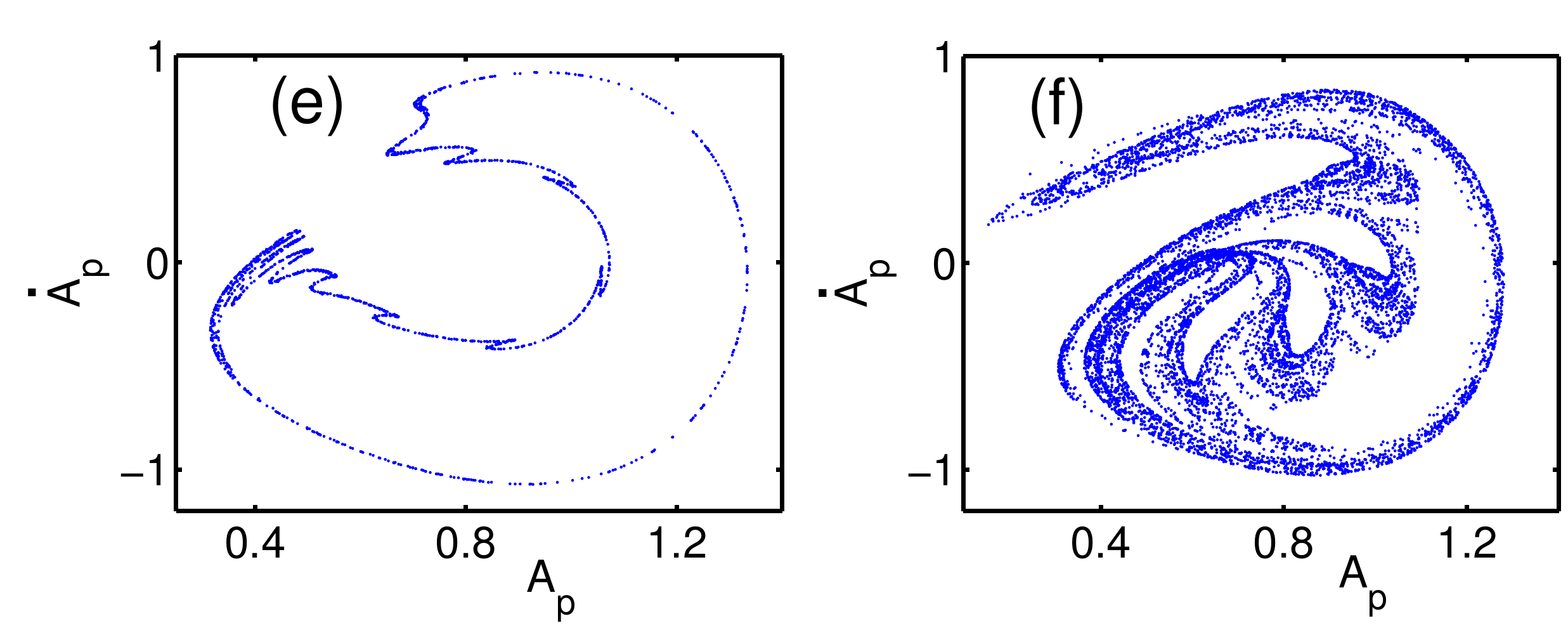}
}
\caption{(a) Quasiperiodic orbit in  the $(\dot A, A, B)$ space for  $h_{ac}=4.4$ keeping $\tau_c = -0.2, \tau_{\epsilon}=-1.0, \Omega =0.97, \gamma =0.0421  ,\xi=3.2, \Gamma'=0.09, h_{dc}= 0.04$ and $ p=0.32$. (b) The corresponding Poincar\'e map in the $(A_p, \dot A_p)$ plane.(c) Poincar\'e map in the $(A_p, \dot A_p)$ plane for $h_{ac}=6.2$ .(d) Largest Lyapunov exponent as a function of $h_{ac}$. (e) Poincar\'e map for a chaotic orbit in the $(A_p, \dot A_p)$  plane for $h_{ac} = 5.7$ and (f) Poincar\'e map for a chaotic orbit in the $(A_p, \dot A_p)$  plane for $h_{ac}=7.38$.}
\label{QP-Route}
\end{figure}

We now examine if our equations also exhibit the quasiperiodic route to chaos as a function of $h_{ac}$ keeping $h_{dc}$ fixed. Here we choose the driving frequency $\Omega =0.97$, but  similar results are obtained for a range of values of $\Omega$ less than unity.  We retain most values of the parameters used for the PD route, but use a higher  coupling $\xi =3.2$ and  a lower damping constant   $\gamma= 0.0421$. In this case, keeping $h_{dc} =0.04$ fixed, we sweep $h_{ac}$ from 2 onwards. We find the first few harmonic frequencies (as found in turbulence of rotating fluid, see Ref. \cite{Gollub75}) in the region $h_{ac} = 2.5$ to $3.8$  beyond which quasiperiodicity is seen till $5.1$ with the emergence of two incommensurate frequencies($\Omega_1=1.86252...$,$\Omega_2=1.79039..$ and another $\Omega_3=\Omega_1-\Omega_2$).  All other dominant frequencies in the power spectrum can be represented by a linear combination of the drive frequency ($\Omega=0.97$) and $\Omega_1$ and $\Omega_2$.  A phase plot of the torus in the $(A, \dot A, B)$ space is shown in Fig. \ref{QP-Route}(a) for $h_{ac}=4.4$. The corresponding Poincar\'e map in the $(A_p,\dot A_p)$ plane is shown in Fig. \ref{QP-Route}(b). The largest Lyapunov exponent vanishes in the regions of quasiperiodicity as can be seen from Fig. \ref{QP-Route}(d). Beyond $h_{ac}=5.1$ till 5.44, we find a phase locked region  arising from a small change in the frequencies participating in the preceding QP regime. In this region of periodicity,  we find four dominant  frequencies all of which are integer multiples ($n=23,24,25$ and $28$) of a new frequency $\Omega_3=0.0746$, which is a rational frequency close to $\Omega_3$ of the QP regime. (The corresponding Poincar\'e map also confirms the region is periodic.) In the interval scanned from $h_{ac}= 2-7.5$,  we find three  regions  of chaos interrupted by windows of quasiperiodicity and periodicity.  The emergence of  the first region of chaos can be attributed to these rational frequencies again becoming incommensurate.   This region of chaos is followed by a transition to a second quasiperiodic regime ($h_{dc} = 5.53 -5.64$) followed by an abrupt transition to  chaos ($h_{dc} = 5.64 - 5.73$). The Poincar\'e map corresponding to quasiperiodic region at $h_{ac}=6.2$, beyond the second region of chaos, is shown in Fig. \ref{QP-Route}(c). Interestingly, the dynamics prior to the first quasiperiodic region (before the emergence of the first region of chaos) and that of after the second region of chaos  has changed considerably as is evident from Fig. \ref{QP-Route}(b) and (c). The third QP and chaotic regions are separated by a periodic regime where we could only detect first five harmonics of the drive frequency since the transition to the third region of chaos is rather sharp. However, the power spectrum in the chaotic regime shows the presence several incommensurate frequencies.   The Poincar\'e maps for the chaotic regions at $h_{ac}=5.7$ and $7.38$ respectively are shown in Fig. \ref{QP-Route}(e) and (f). The quasiperiodic  route to chaos is observed for a range of values of parameters around the values used for Fig. \ref{QP-Route}.

\begin{figure}
\vbox{
\includegraphics[height=10.6cm,width=7.5cm]{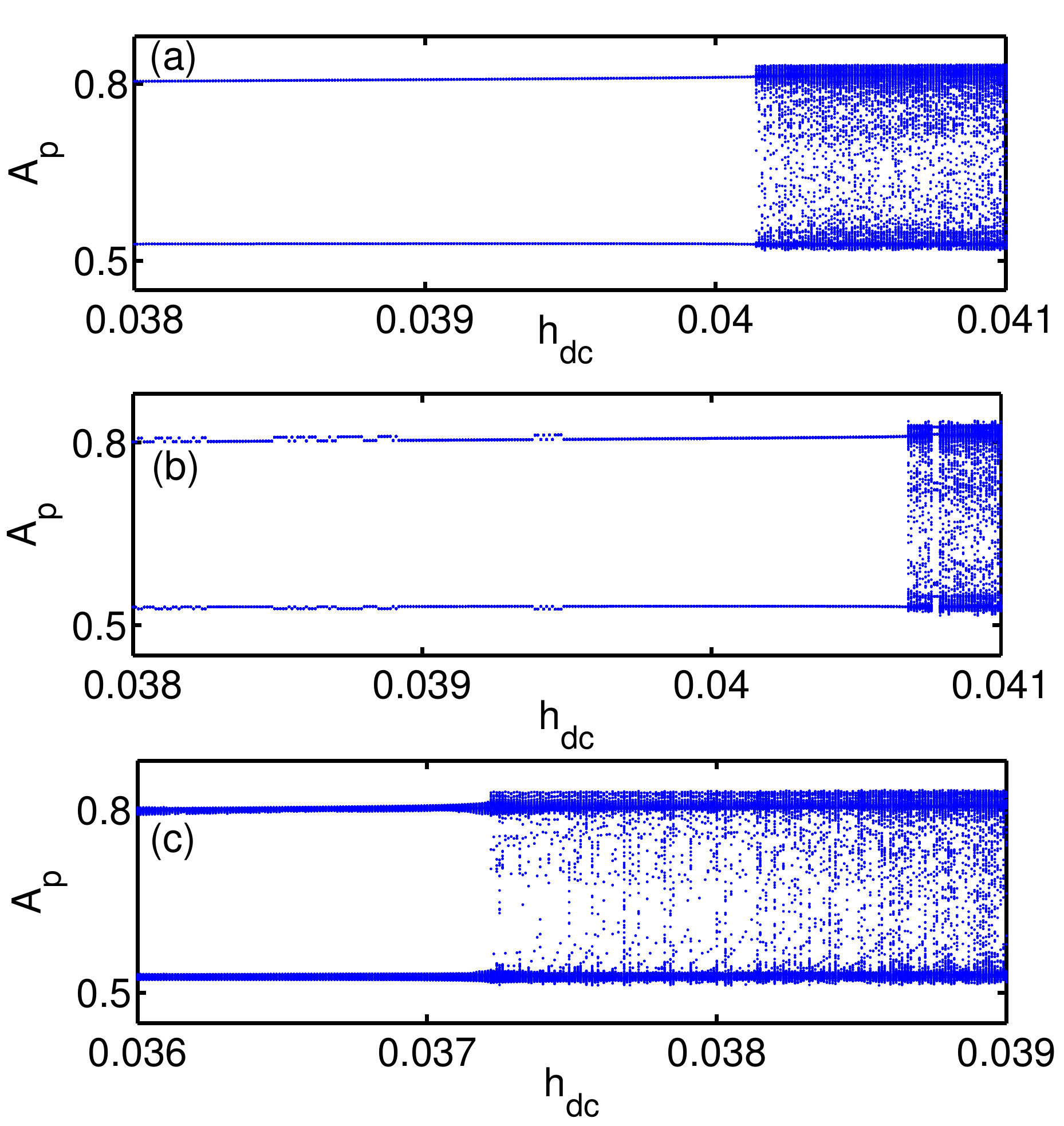}
\includegraphics[height=3.55cm,width=7.0cm]{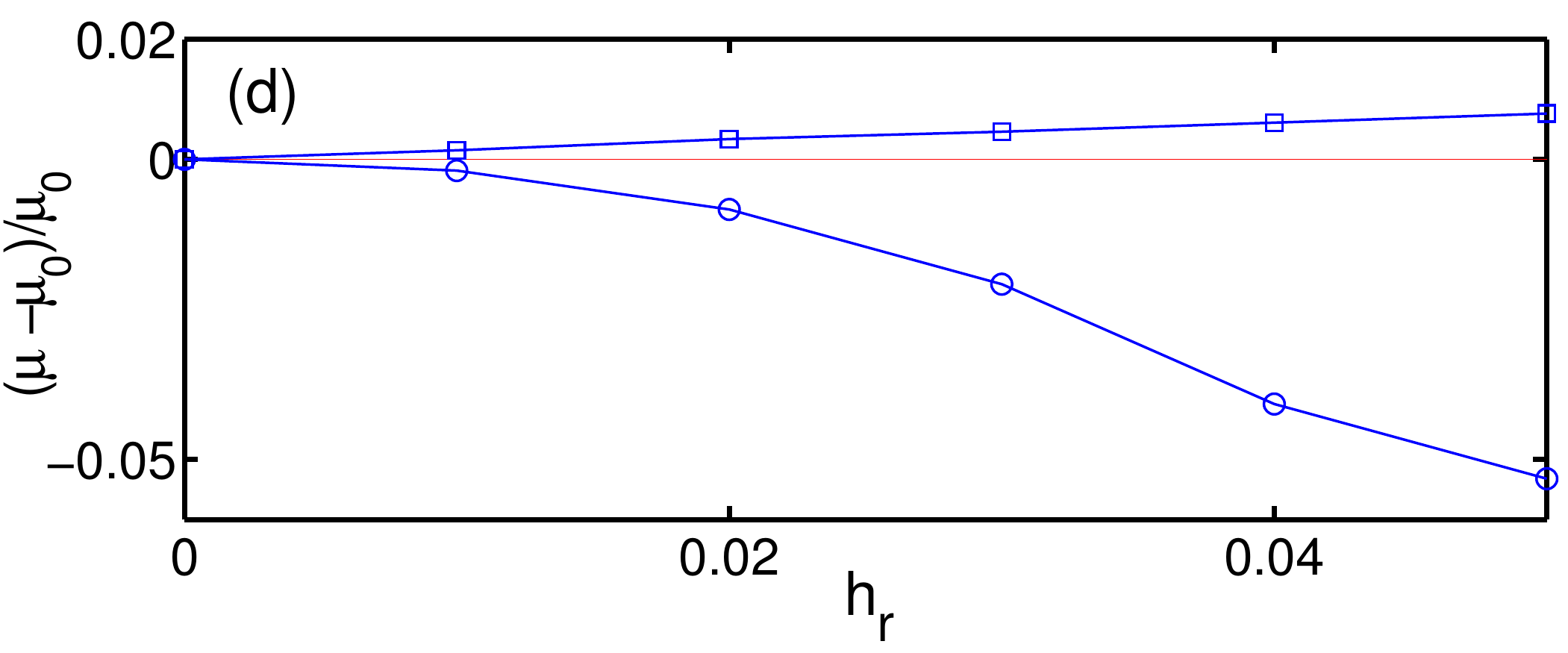}
}
\caption{(a) Bifurcation diagram as a function of $h_{dc}$ for $\tau_c = -0.2, \tau_{\epsilon}=-1.0,\xi=1.0, \Omega =1, \gamma= 1.592,\Gamma'=0.09, h_{ac}= 10.5$ and  $p=0.32$.  (b) Bifurcation diagram as a function of $h_{dc}$ for the same set of parameters for  $\delta =10^{-6}$ and $h_r=0.05$ that clearly shows  delayed chaos. Note that the  period two orbit appears to switch by a small amount intermittently. (c) Bifurcation diagram as a function of $h_{dc}$ for the same set of parameters for  $\delta =10^{-2}$ and $h_r=0.05$ that clearly shows  induced chaos.  (d) Suppressed and induced chaos for $\delta =10^{-6}(\square)$  and $\delta=10^{-2}(\circ)$ as a function of $h_r$.  }
\label{Shift-Chaos}
\end{figure}

We next investigate if our model predicts  induced and suppressed chaos reported in experiments when the system is subjected to an additional small amplitude near resonant sinusoidal perturbation \cite{Vohra95}. A number of theoretical and experimental studies devoted to nonfeedback control of chaos use  several variants of a weak harmonic perturbation such as   parametric perturbation of suitable frequency \cite{Lima90}, an additive  forcing  \cite{Braiman91, Qu95} and  resorting to change in the shape of additive signal etc \cite{Chacon93}. Some studies emphasize the importance of the phase factor as well. Fewer studies have been reported in the near resonance  condition \cite{Chiz97} as in the experiments of Vohra {\it et al} \cite{Vohra95}.  However in the experiments, the authors  make no reference to the phase difference between the driving ac field and the resonant drive. For this reason we use a resonant drive  of the form $h(\tau) = h_r \sin \Omega_r \tau$, with $\Omega_r$ close to the first subharmonic $\Omega_r = \Omega/2 - \delta$ (or the second subharmonic).    In their experiments, the influence of the additional resonant field has been illustrated with respect to a transition from period-$2$ orbit to chaos as a function of $h_{dc}$ \cite{Vohra95}. Thus, we first locate a direct transition from period two cycle to chaos as a function of $h_{dc}$ keeping the perturbing field $h_r=0$. This transition occurs at $h_{dc} = 0.04014$. (The parameter values  used are the same as for the period doubling  route except for $\xi=1.0$.)
A bifurcation diagram for  the strobed strain amplitude $A_p$ is shown in Fig. \ref{Shift-Chaos}(a). First we consider keeping  $\delta$ small. Keeping $h_{dc}=0.04014$, we apply the perturbing signal $h_r \sin \Omega_r \tau$ with $\delta = 10^{-6}$. Even for small $h_r$ we find that the onset of chaos is delayed as is clear from Fig. \ref{Shift-Chaos}(b). Another feature that is clear (from Fig. \ref{Shift-Chaos}b) is that the amplitude of the period two orbit appears to switch between values corresponding to the absence of resonant field and a slightly increased value in the presence of the near resonant field. The switching between the two values is also  intermittent with $h_r$. Other changes  can also be noted. For example, just beyond the delayed onset of chaos, we see a period five window. Thus, not only there is delayed chaos, there are some changes in the bifurcation diagram as well. The magnitude of the shift, though small, increases linearly with $h_r$. 

We have also studied the influence of $\delta$ on the nature of the dynamics as in experiments. For illustration we use $\delta = 10^{-2}$. In this case, on examining the  amplitude $A$ as a function of time, we find that  the effect of the resonant  sinusoidal drive manifests as a periodic modulation of the amplitude of the period two orbit with a period $T= 2\pi/\delta$ ($ = 2 \pi \times 10^2$ for $\delta=10^{-2}$).  The amplitude of the modulation is proportional to $h_r$.  We find that these modulations persist even after long time ($1-3\times 10^7$ time steps of $\Delta t=0.01$).  Thus, the associated Poincar\'e map of the periodic orbit shows finite dispersion.  Even so, a simple visual inspection of the bifurcation diagram displayed in Fig. \ref{Shift-Chaos}(c) shows that the onset of chaos is hastened. This is confirmed by the Lypaunov exponent calculation. (It must be stressed here that even for  small  $\delta$, such a periodic  modulation of  the amplitude of the periodic two orbit is seen. However, since the  period of this modulation is large $= 2\pi \times 10^6$ and since the Poincar\'e map is constructed over time duration small compared to the period of this modulation,  the dispersion in the Poincar\'e map is not visible for $\delta = 10^{-6}$.) One can also note that the magnitude of the shift in the onset of chaos ($\circ$) is significantly larger than that  for the suppressed chaos as can be seen from  Fig. \ref{Shift-Chaos}(d).   We also note that  in our case, suppressed chaos is seen for small values of $\delta$ while induced chaos is seen for large values of $\delta$, opposite of what is reported. 

Identifying $\mu_0$  with the onset of chaos ($h_{dc} =0.04014$) when $h_r =0$  and $\mu$ that for finite $h_r$,  the normalized shift ($\square$) $\frac{\mu-\mu_0}{\mu_0}$ for the onset of chaos (obtained by computing  the Lyapunov exponent) as function of $h_r$  is shown in Fig. \ref{Shift-Chaos}(d). Using $ \Omega_r= \Omega/2 + \delta$ does not affect induced or delayed chaos, but the magnitude of the shift is different.

\begin{figure}
\vbox{
\includegraphics[height=3.5cm,width=7.0cm]{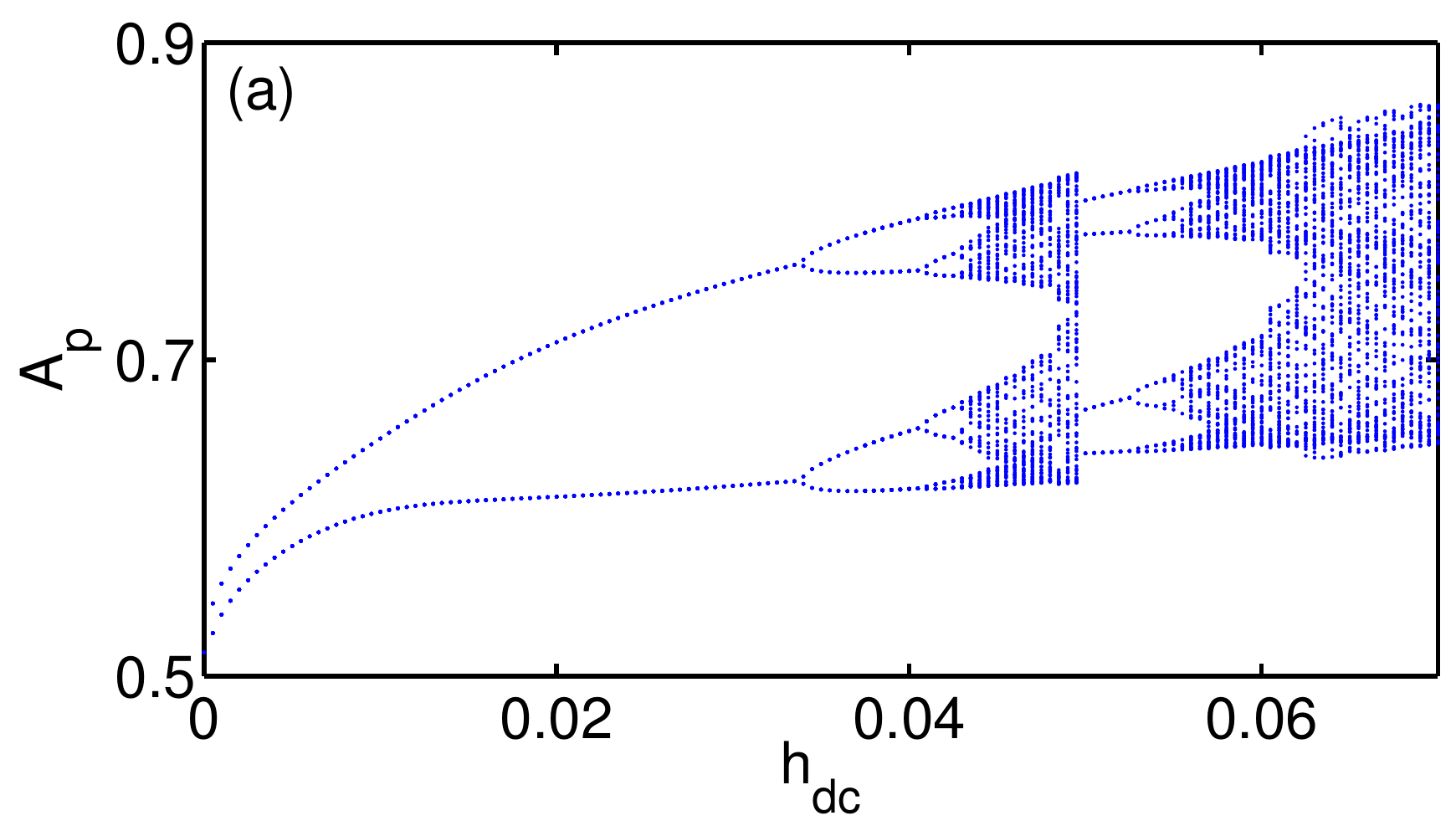}
\includegraphics[height=3.5cm,width=7.0cm]{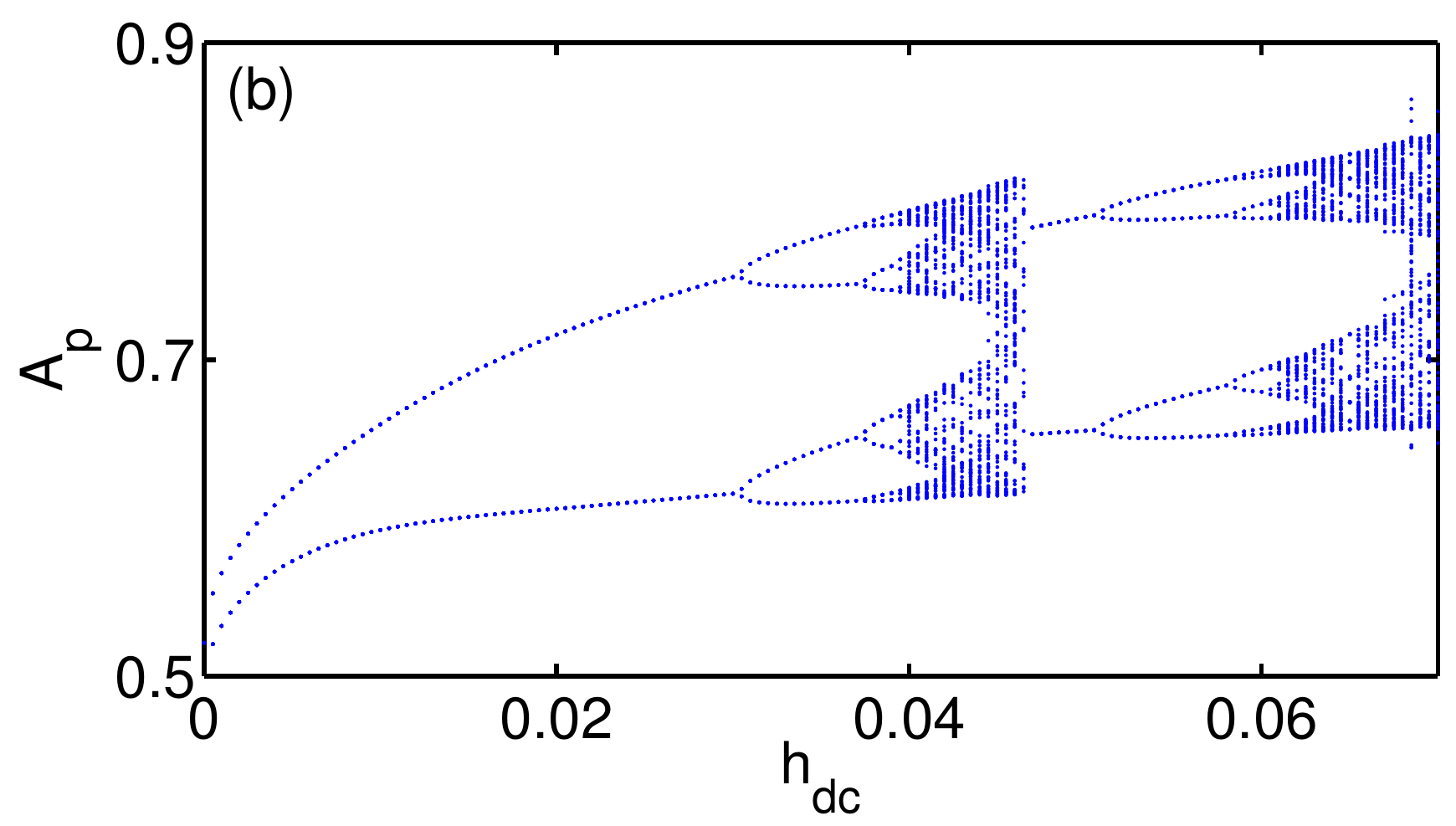}
}
\caption{(a) Bifurcation diagram as a function of $h_{dc}$ for $\tau_c = -0.2, \tau_{\epsilon}=-1.0,\xi=0.6, \Omega =1, \gamma= 1.592, \Gamma'=0.09, h_{ac}= 10.5, p=0.32 $ and $h_r=0.02$.  (b) Bifurcation diagram as a function of $h_{dc}$ for $\tau_c = -0.2, \tau_{\epsilon}=-1.0,\xi=0.6, \Omega =1, \gamma= 1.592, \Gamma'=0.09, h_{ac}= 10.5, p=0.32$ and $h_r=0.035$.  }
\label{Forcing-PD-Bif}
\end{figure}

We now examine how  the near resonant forcing affects  the period doubling sequence as a function $h_{dc}$ shown in Fig. \ref{PD-Bif}a.  (The parameter values used are the same as for the PD route shown in Fig. \ref{PD-Bif}).  Consider  resonant forcing with respect to the first subharmonic, ie., $\Omega_r = \Omega/2 - \delta$ keeping  $\delta = 10^{-6}$.  In the absence of $h_r$, while the PD route starts  with period one cycle (see Fig. \ref{PD-Bif}a), even for very small resonant forcing amplitude, say  $h_r =0.02$, we do not find the period one cycle.  A plot of the bifurcation diagram keeping $h_r=0.02$ is shown in Fig. \ref{Forcing-PD-Bif}(a).  As can be seen from the figure, not only the bifurcation digram starts with period two cycle but also the entire bifurcation diagram is affected (compare Fig. \ref{PD-Bif}(a)). While successive PD bifurcations  leading to chaos are hastened, a period four orbit interrupts the chaotic band. As we increase $h_{dc}$, the bifurcation sequence  also  culminates in chaos. For larger values of $h_r$(say $0.035$), the first chaotic band is interrupted by a period two orbit instead of period four orbit seen for smaller $h_r$. This is shown in Fig. \ref{Forcing-PD-Bif}(b). This feature of altering the entire bifurcation diagram in the presence of a resonant perturbing field is similar to earlier studies on the effect of nonresonant perturbing sinusoidal field \cite{Chiz97}.

However, using larger $\delta$, say $\delta = 10^{-3}$, even for small $h_r=0.02$,  transients persist even after very long runs ($10^7$ time steps). The influence on  period two cycle is so significant that one sees a band in the Poincar\'e map instead of two points. This is due to fact that the sinusoidal modulations of period $2 \pi/\delta$ with amplitude proportional to $h_r$ persist even after long runs. The magnitude of  fluctuations become  significant near bifurcation points. We have verified that a similar effect is also seen in the case of the Duffing's Oscillator.

\begin{figure}[t]
\vbox{
\includegraphics[height=4.0cm,width=7.5cm]{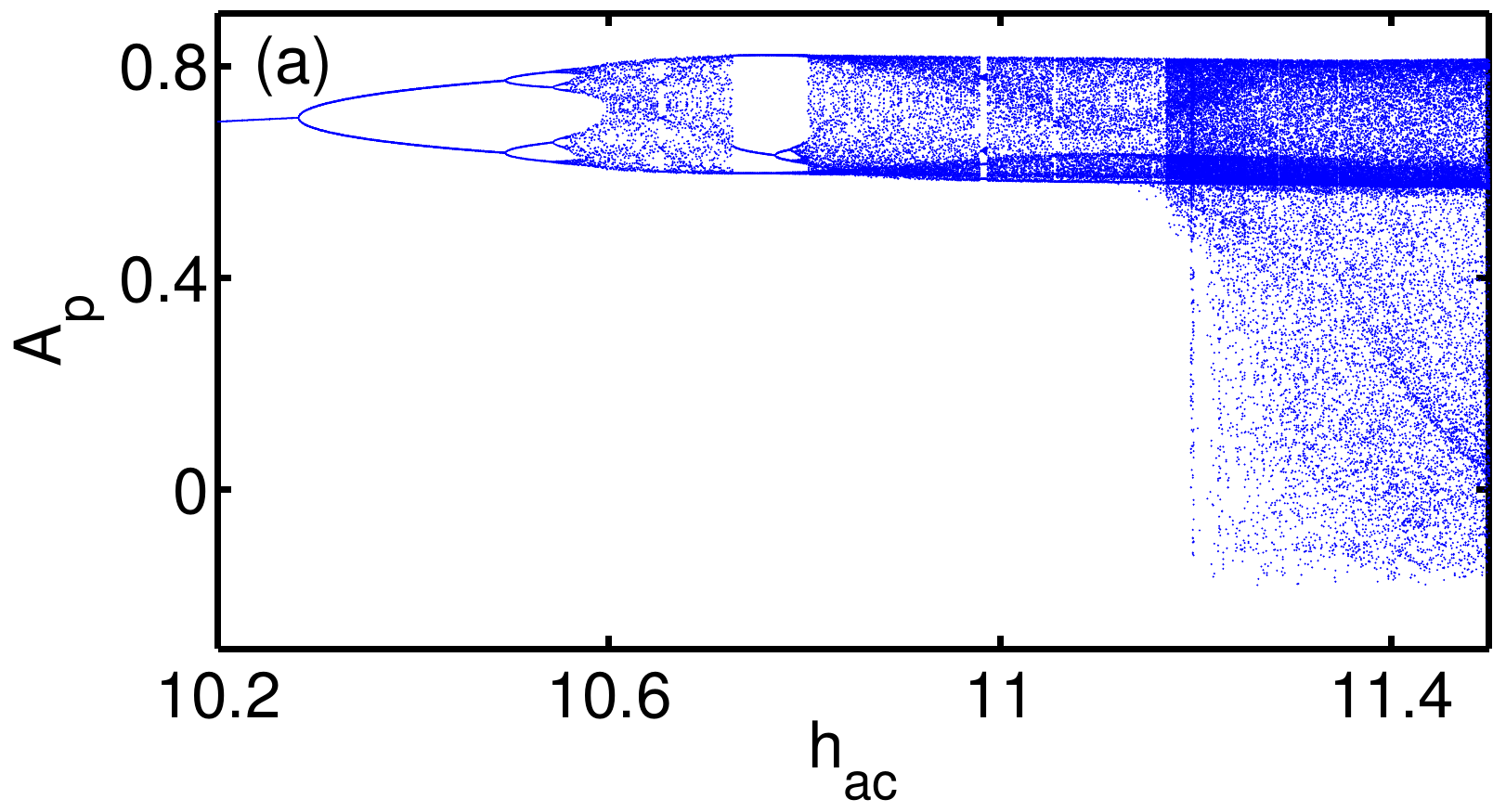}
\includegraphics[height=4.0cm,width=7.5cm]{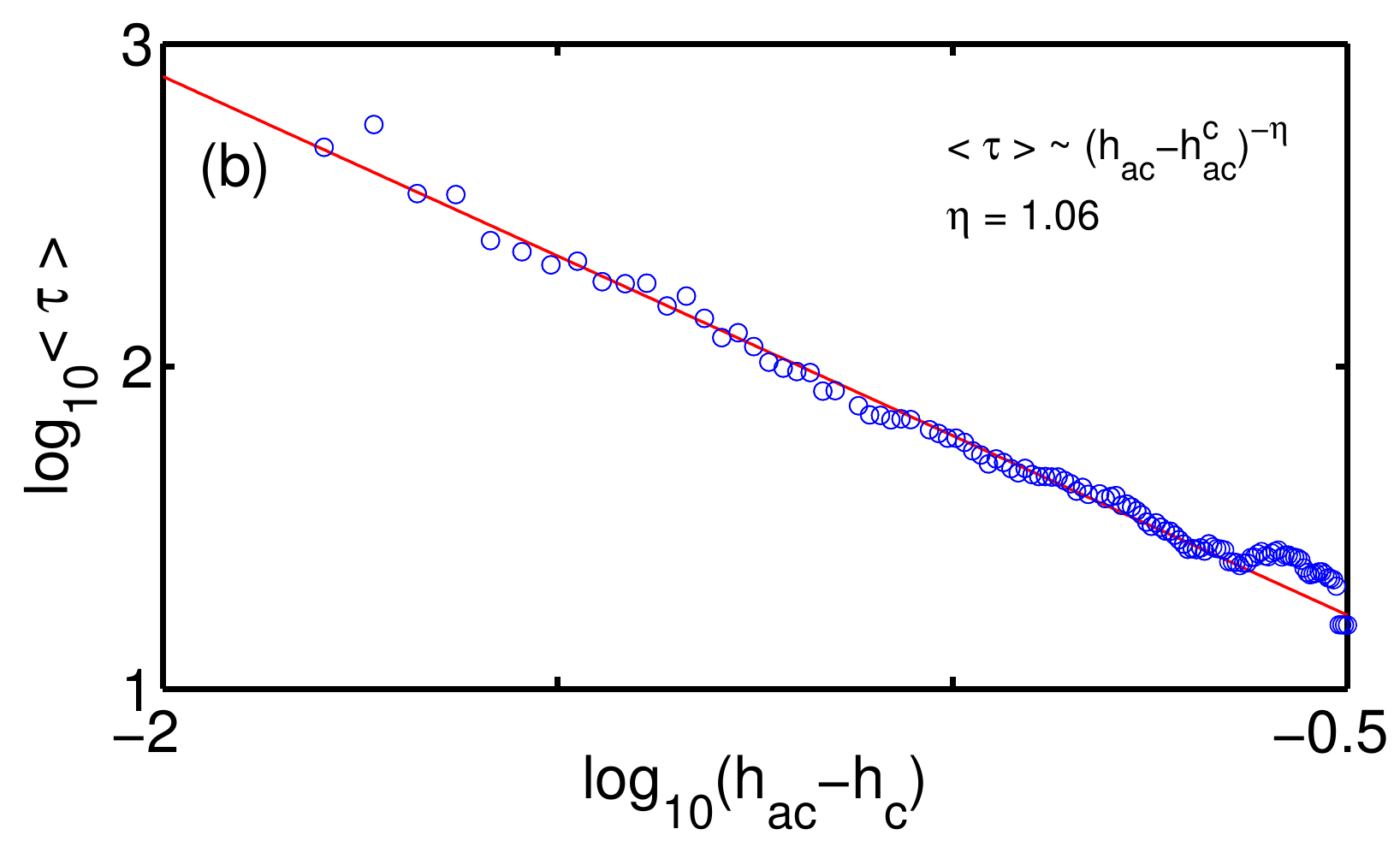}
}
\caption{(a) Symmetry restoring crisis during period doubling bifurcation as a function of $h_{ac}$. The parameter values are the same as used in Fig. \ref{PD-Bif}, except that we keep $h_{dc}=0.04$ and $\beta =1.4$ and vary $h_{ac}$. Other parameter values are $\tau_c = -0.2, \tau_{\epsilon}=-1.0,\Omega=1,\xi=0.6, \gamma= 1.592, \Gamma'=0.09$ and  $p=0.32$.  (b) Power law dependence of the mean residence time in the pre-attractor. 
}
\label{Sym-Res-Crisis}
\end{figure}

As is well known, driven oscillators exhibit several interesting dynamical features such as persistent chaotic transients noted near crisis points \cite{Grebogi82, Grebogi87}. For example, for a system with two wells, symmetry restoring crisis is usually seen \cite{Ishii86, Grebogi87}. Thus, symmetry restoring crisis is a distinct possibility since our model has left-right symmetry. Here we investigate this possibility in the case of the  period doubling route to chaos observed as a function of $h_{ac}$ keeping $h_{dc}$ fixed at a suitable value. For a given set of initial conditions located in one well, the attractor is confined to the same well for a range of values of $h_{ac}$ starting from zero.  However, as  we  increase $h_{ac}$ further the attractor suddenly expands to occupy the symmetry allowed phase space corresponding to the two wells  at a critical value  $h_{ac}= h_{ac}^c = 11.194$. A plot of the bifurcation diagram showing the sudden expansion of the attractor is displayed in Fig. \ref{Sym-Res-Crisis}(a).  Note also that near $h_{ac}^c$, the attractor expands slowly to occupy larger positive values of the original well before jumping to the left well. Very long runs were required  near the transition regime. To  find transient chaos as we approach the critical value from above, we have numerically calculated the mean residence time $<\tau>$ in the pre-crisis attractor for various values of $h_{ac}$. We find that $<\tau>$ scales as a power of $h_{ac} - h_{ac}^c$, ie., $ <\tau> \sim (h_{ac} - h_{ac}^c)^ {- \eta}$ where $\eta$ is an exponent \cite{Ishii86, Grebogi87}. A log-log plot of the mean residence time in the pre-crisis attractor is shown in Fig. \ref{Sym-Res-Crisis}(b). The value of the exponent obtained turns out to be $\eta \sim 1.06 $. 

\subsection{Driven magnetoelastic beam}

The above model equations can be easily adopted to describe the experimental results of the dynamics of a magnetoelastic beam  \cite{Moon, Moon84, Cusumano95}. Here, we target experiments where both the phase plots and the associated Poincar\'e maps have been reported  \cite{Cusumano95}.   These authors first obtain the static bifurcations of the beam exhibiting two equivalent symmetric strains either through subcritical or supercritical bifurcation depending on the distance of the drive from the epoxied magnet. Consider the case of supercritical static bifurcation which implies  $\tau_{\epsilon} < 0$ in our model. Further, the material is a crystalline alloy which corresponds to $p=1$ in our model.  The frequency of the drive used in their study is much less than the natural frequency of the beam. Further the damping is stated to be low which implies that we should use small $\gamma$ value in our model. The authors monitor the phase plot in the $(A, \dot A)$ plane and  also obtain the corresponding Poincar\'e map. Our purpose is to see if these two results can be reproduced  by using  our model equations. 

\begin{figure}[t]
\vbox{
\includegraphics[height=4.0cm,width=7.5cm]{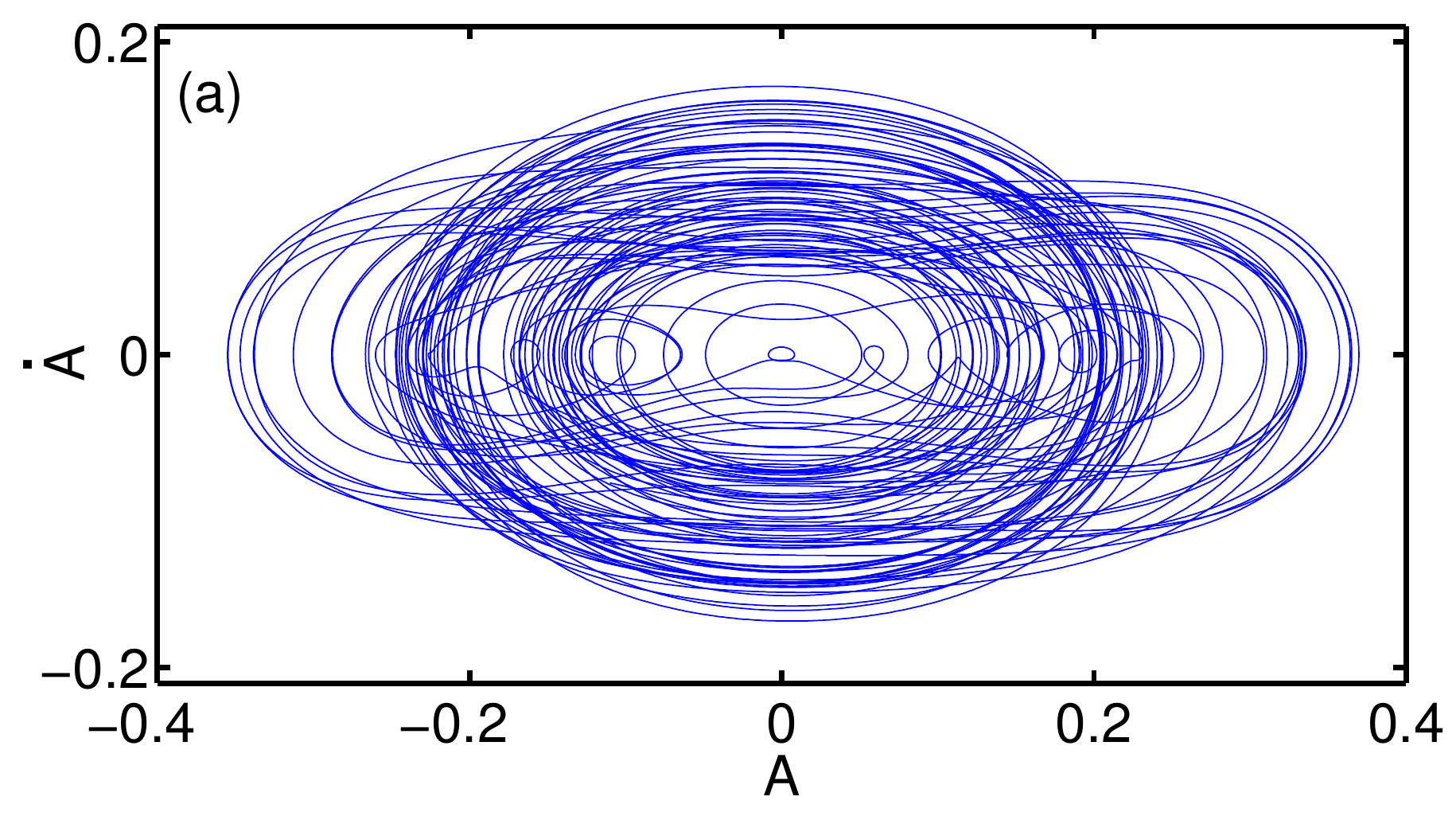}
\includegraphics[height=4.0cm,width=7.5cm]{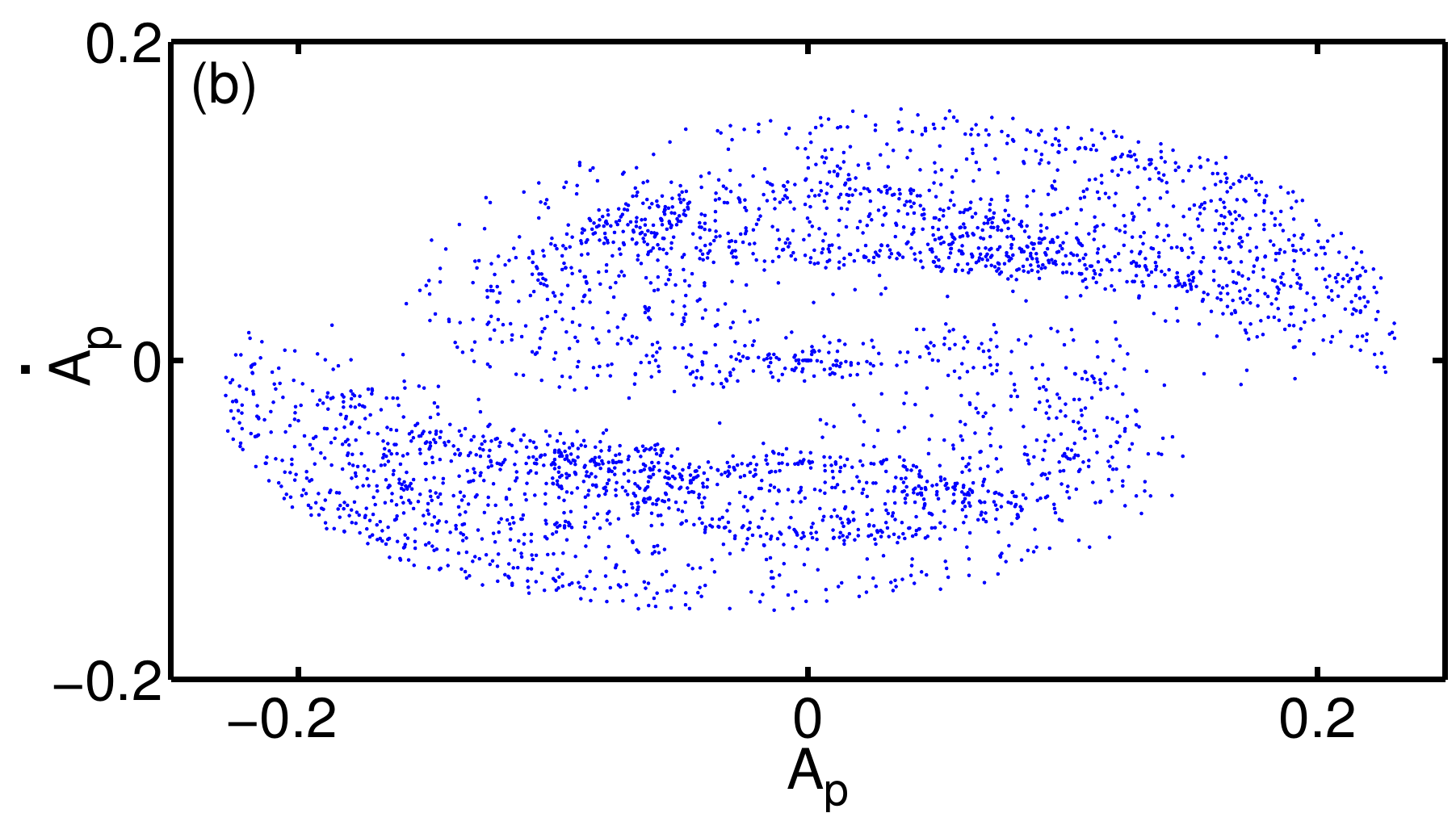}
}
\caption{(a) Phase plot in the $(A, \dot A)$ plane for  $\tau_c = -0.1, \tau_{\epsilon}=-0.25, \Omega=0.13,\xi=3.7, \gamma= 0.05, \Gamma'=0.1, p=1.0, h_{dc}=0, \beta=\Delta =1$ and  $h_0=0.42$.  (b) The corresponding Poincar\'e plot.
}
\label{Mag-Elas}
\end{figure}

We have numerically solved Eqs. (\ref{A}, \ref{B}) keeping the parameters fixed at  $\tau_c = -0.1, \tau_{\epsilon}=-0.25, \Omega=0.13,\xi=3.7, \gamma= 0.05, \Gamma'=0.1, h_{dc}=0, \beta =1$, and $ \Delta =1 $. Since experiments use low frequency, for illustration we use $\Omega =0.13$, which is nearly a factor ten smaller than the natural frequency of the system($\Omega =1$). Initially for small $h_{ac}$ we find that the strain amplitude decays. As we increase $h_{ac}$ beyond a critical value,  we see a   $(A,\dot A)$ phase plot similar to Fig. 5(a)  of Ref. \cite{Cusumano95}. However, in our case, since the phase plot is a projection of the three dimensional flow (corresponding to elastic and magnetic degrees of freedom) on the $(A, \dot A)$ plane, the Poincar\'e map can be sensitive to the value of $h_{ac}$. Varying $h_{ac}$, we find  a Poincar\'e map similar to Fig. 5(b) of Ref. \cite{Cusumano95} for $h_{ac}=0.42$.  The phase plot in $(A, \dot A)$ plane and the associated Poincar\'e map are shown in Fig. \ref{Mag-Elas}(a) and (b) respectively.  Clearly these plots are very similar to those in  Ref. \cite{Cusumano95}. Similar Poincar\'e maps (as in experiments) are seen for a range of values beyond $h_{ac}= 0.38$.

\section{Driven martensites}

As stated in the introduction, the model equations can also be adopted to describe the dynamics of nonmagnetic martensitic ($Cu_{82.9} Al_{14.1}Ni_3$) ribbon carried out in the context of internal friction experiments quite early \cite{Suzuki80, Wuttig81}. These experiments have been performed by fixing the sample at one end and applying a sinusoidal drive at the other end. Multiperiodic oscillations have been reported.  Before proceeding further, we note that it is traditional to use a sixth order Landau polynomial in $\epsilon$ such that the martensite minima occur at $\epsilon = \pm 1$. This is equivalent to using  $\beta = 2$ and $\Delta = 3$ in the  free energy.  Further, we note that the sample has a low $Ni$ content, which however, is adequate to enforce periodic oscillations using an applied magnetic field $h_{ac}$. Alternately, small permanent magnet is attached that help to drive the free end of the sample \cite{Wuttig81}. Since the alloy is nonmagnetic, we can drop the equation for $m$ and terms containing $B$ in Eq. (\ref{A}). Then we have an equation for the strain variable that reduces to the form derived by Bales and Gooding \cite{Bales} for a martensite in one dimension.  We note that the equation for the strain variable with the dissipative term gives twinned solutions (near square wave type of solutions, see Fig. 1 of Ref. \cite{Bales}). Thus, we have a single equation for the strain amplitude with a driving term $h_{ac} sin \Omega t$ given by 
\begin{eqnarray}
\nonumber
\ddot{A}(\tau)&=&-\Big[\big(\tau_{\epsilon} +k^2\big) A  +6  A^3(\tau)-\frac{15}{8}A^5(\tau) +\gamma k\dot{A}(\tau)\\
\label{MART}
&&-h_{ac} sin \Omega t\Big].
\end{eqnarray}
Here, $\tau_{\epsilon} = \frac{T-T_s}{T_m-T_s}$, where $T_m$ is the martensite onset temperature.  Note that the effective transformation temperature is $\tau_{\epsilon} + k^2$.

\begin{figure}[t]
\vbox{
\includegraphics[height=4.0cm,width=8.0cm]{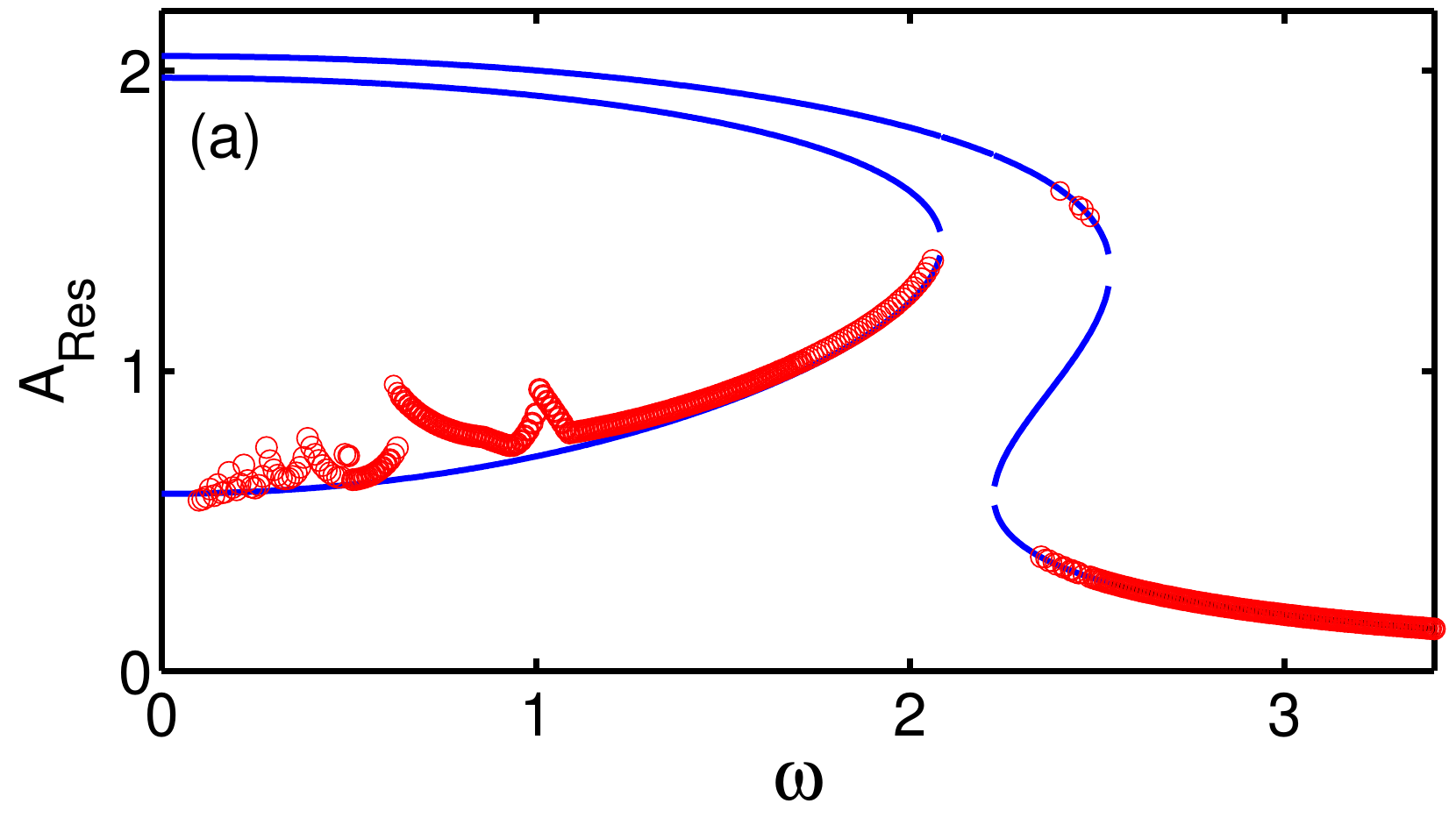}
\includegraphics[height=4.0cm,width=8.0cm]{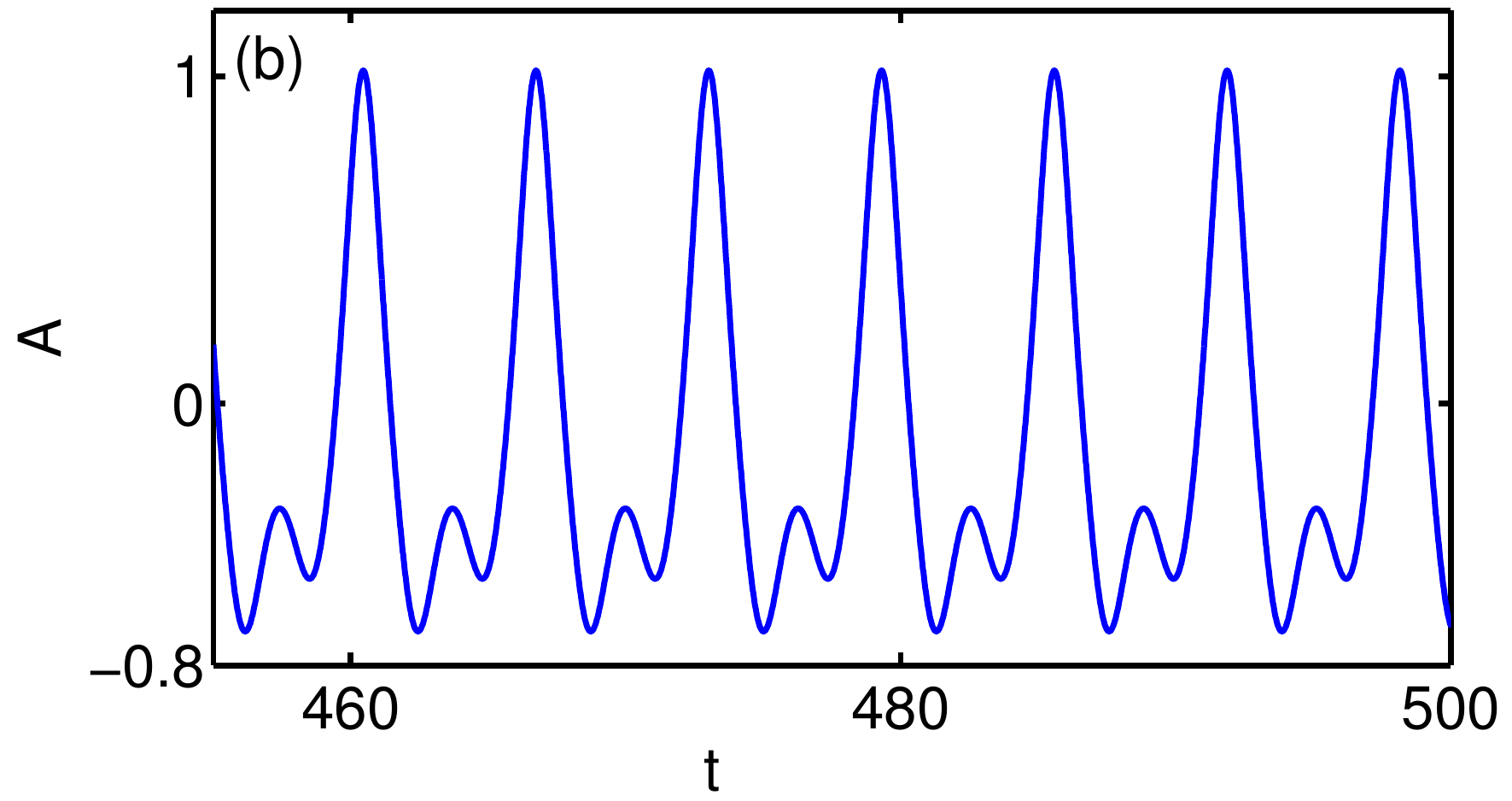}
}
\caption{(a) Resonance curve given by Eq. (\ref{ResonanceE}) for $h_0=1.46, \gamma=0.25$ and $\tau_{\epsilon}=0.95$. Also shown is the numerically obtained resonance curve that shows several resonances for $\Omega$ less than unity. (b) A typical amplitude plot for $\Omega = 0.99995$. 
}
\label{Resonance}
\end{figure}

First we address the presence of multiperioidic oscillations when the read is driven at  near resonant frequency  keeping the temperature just above the  martensite transformation temperature $T_m$. The resonance curve corresponding to Eq. (\ref{MART}) is given by 
\begin{equation}
\Big[  ((\tau_{\epsilon} + k^2)-\Omega^2)  A + \frac{9A^3}{2} -\frac{75A^5}{64} \Big]^2 + \big(\gamma k\Omega A\big)^2 = h_{ac}^2.
\label{ResonanceE}
\end{equation}
A plot of the resonance curve is shown in Fig. \ref{Resonance} for driving amplitude $h_{ac}=1.46$. Clearly there are regions of multistability that are realized when the initial conditions are in suitable  basins of attraction.   The gap between the upper two lines closes  for small values of $h_{ac}$.  Also shown in the plot are the numerically obtained resonance curve marked by open circles that exhibit several resonances for $\Omega \le 1$.

First,  we briefly summarize the results of the internal friction experiments \cite{Suzuki80, Wuttig81}.  A cursory look at the figure of relaxational oscillations (Fig. $2$ of Ref. \cite{Suzuki80}) shows that there are four distinct peaks with a rough periodicity of $310\, s$. (The nature of the curve for low amplitudes is considerably rough which can be attributed to lack of control in measuring devises or instrumental errors.) The authors state that these oscillations are seen beyond a certain amplitude of the drive.  While the authors recognize that these are nonlinear oscillations, they fail to recognize that they are subharmonics since they specifically state that they are harmonics. (However, it is not even clear if these are stabilized oscillations.) Similarly, the plots of the strain amplitude given in Figure $3$ and $4$ of Ref.\cite{Wuttig81} also appear to be multiperiodic. In the absence of reliable plots of these relaxational oscillations we regard them as multiperiodic.  

We now examine the response of Eq. (\ref{MART}) under conditions close to those in experiments. These experiments have been conducted at a temperature close to martensite transformation temperature ($1^0$ above $T_m$) and under resonant forcing conditions.  The natural frequency of the ribbon in our model is unity.  We use a drive frequency $\Omega = 0.99995$ (that roughly corresponds to $\delta \omega =0.005 s^{-1}$ when the natural frequency is $\sim 100$). Using  $\tau_{\epsilon} =0.95$,  the effective temperature  $\tau_{\epsilon} + k^2 = 1.05$ corresponding to a temperature to just above $T_m$.  When the amplitude of the drive $h_{ac}$ is small, we see a wave form that follows that of  the drive. This is related to the small  gap between the two upper branches which opens-up as we increase $h_{ac}$. This happens around $h_{ac} =1.35$. For values beyond this  we see a period two cycle. A typical period two cycle for $h_{ac} \sim 1.46$ and $\Omega=0.99995$ is shown in  Fig. \ref{Resonance}(b).  As can be seen, the wave form is qualitatively similar to the experimental plot even though they are of period two. (See Fig. 2 of Ref. \cite{Suzuki80}.) This also corresponds to the resonance near $\Omega \sim 1$. We also find period three, five and seven cycles at the resonances found with decreasing $\Omega$. However, we do not see  period four.

\begin{figure}[t]
\vbox{
\includegraphics[height=4.0cm,width=7.5cm]{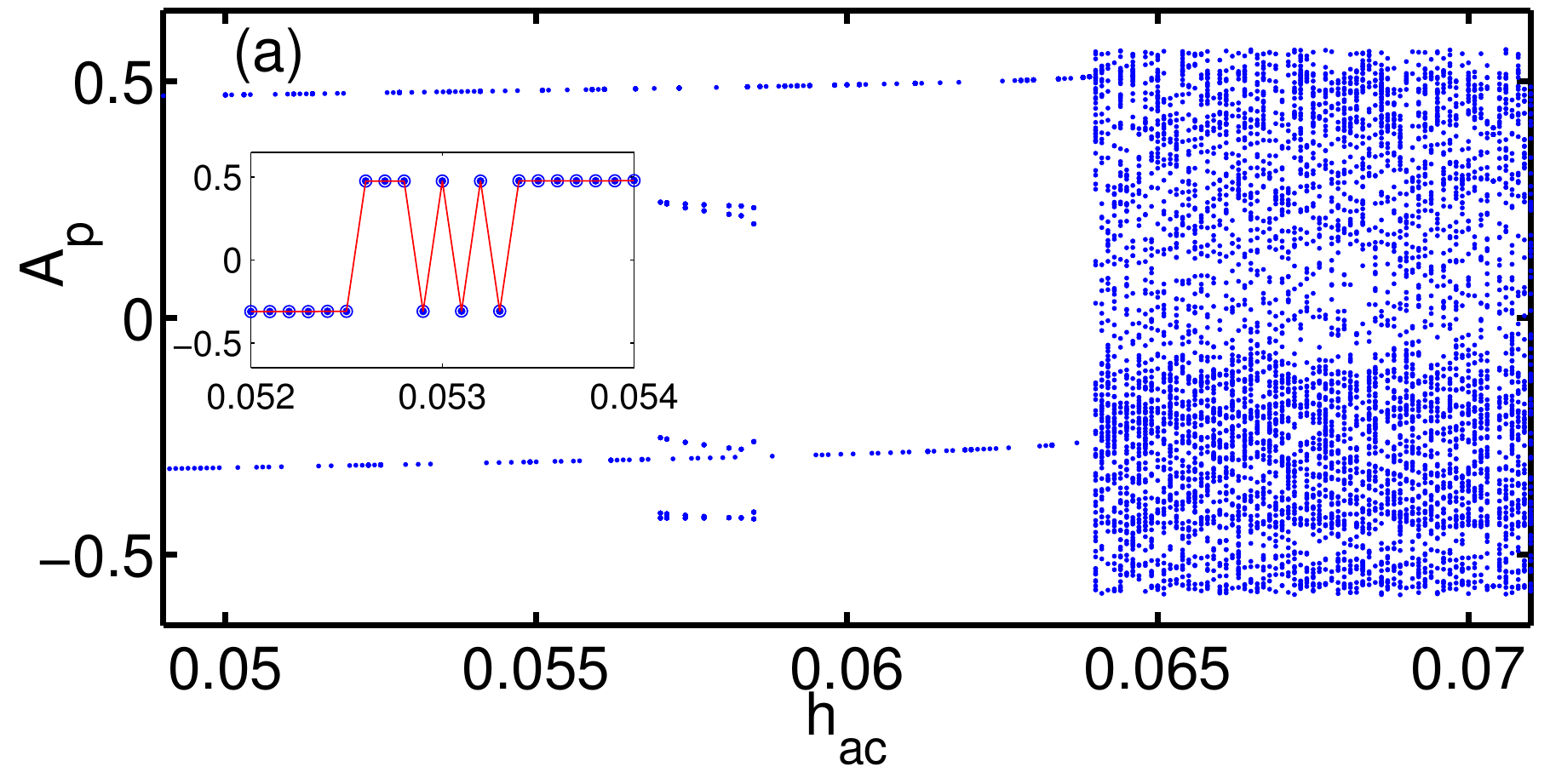}
\includegraphics[height=4.0cm,width=7.5cm]{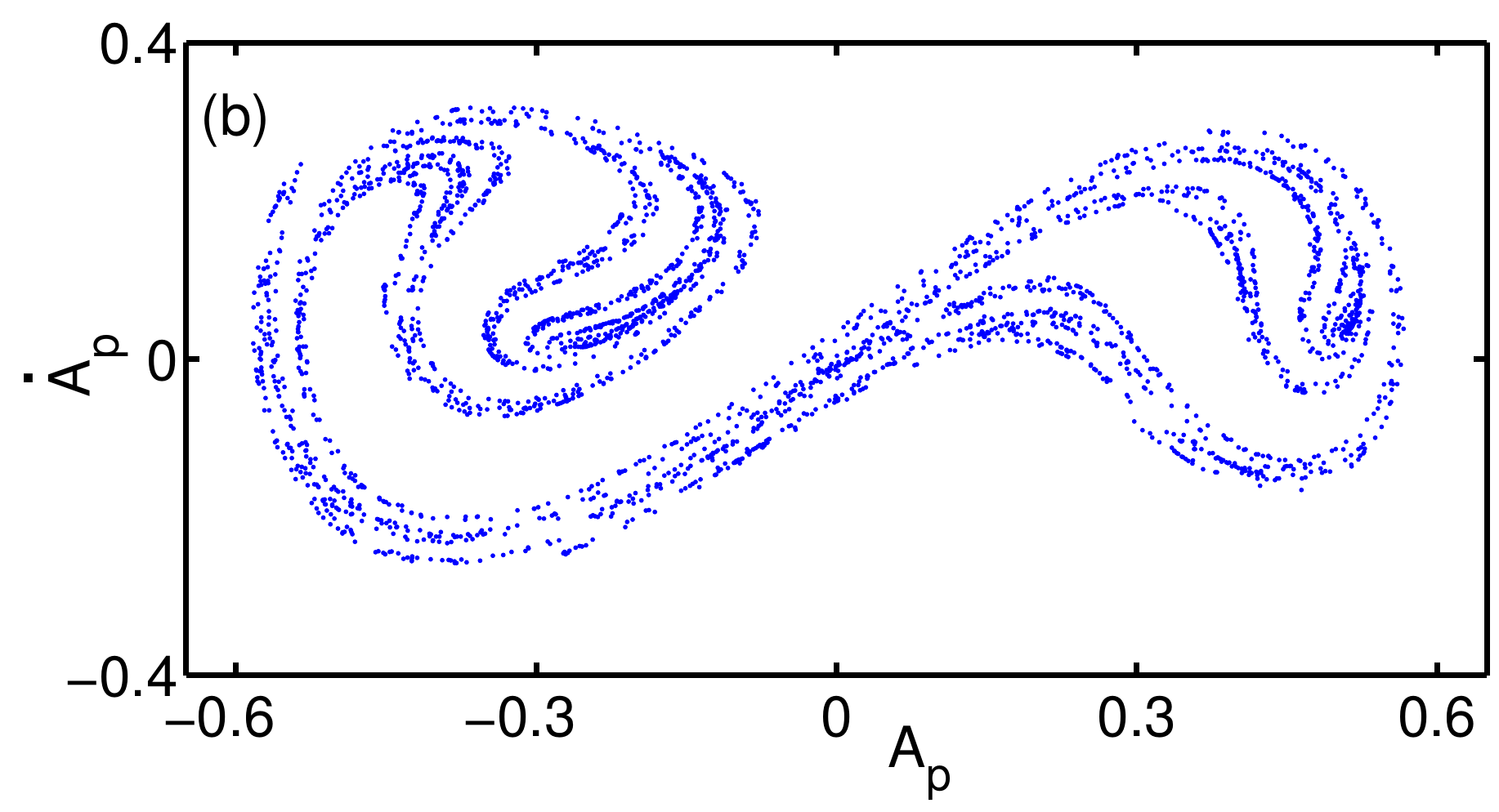}

}
\caption{(a) Bifurcation diagram for $\tau_{\epsilon}=-1,\Omega=0.99995,\beta=2; \Delta =3; \gamma=0.25$  (b) A typical  Poincar\'e map in the chaotic regime for $h_{ac}=0.64$. 
}
\label{Switching-Dyn}
\end{figure}

The dynamics  is more interesting  when the temperature is lower than $T_M$.  Usually resonant driving conditions are expected to exhibit only periodic response. However, in this case, we find unusual dynamics. To illustrate this, we choose $\tau_{\epsilon}= -1$ with $\Omega = 0.99995$.  For small values of $h_{ac}$, with initial conditions in one of the wells,  the strain amplitude remains confined to the well. Beyond a certain value of $h_{ac}$,  typically the barrier height of the potential well ($\sim 0.035$ for these parameter values), the system still settles to a periodic state confined to the same well. As we increase the drive amplitude $h_{ac}$ further,  the transients become long. It is beyond this value that we find that the orbit can settle into either of the two wells starting from the same initial condition  in one of the wells.  The switching dynamics is seen when  $h_{ac}$ exceeds $0.038$. A bifurcation diagram of the strobed amplitude $A_p$ is shown in Fig. \ref{Switching-Dyn}a. The inset clearly shows  the dynamics of the period one cycle switching between the two wells.  We could not find any regularity in this switching process as a function $h_{ac}$ (see the inset).  As can be seen from Fig. \ref{Switching-Dyn}a,  we find period five and period one are interspersed in the region $0.057 < h_{ac} < 058$. Intriguingly, we could not find any pattern in the duration of these transients as we increased $h_{ac}$. Sometimes even a very small change in $h_{ac}$ led to very short transients and at other times it resulted in long transients. Another interesting feature is that for certain values of $h_{ac}$,  we find long transients that appeared  to be period five,  which eventually settled to period one cycle confined to one of the wells. The switching dynamics is seen till $h_{ac}= 0.646$. Further increase in $h_{ac}$ leads to  a direct transition from the periodic one orbit to chaos  as shown in Fig. \ref{Switching-Dyn}. We note here that  a direct transition from a period one cycle to chaos has been realized in the study of Josephson junctions under nonresonant drive conditions \cite{Nayak07}. Moreover, the amplitude of the periodic one orbit does not fluctuate. In contrast, in our case, the dynamics is observed under near resonant drive conditions. To the best of our knowledge, we are not aware of any reports on this kind of switching phenomenon. More work is needed to understand the underlying mechanism leading to the switching dynamics.

\section{Discussion}

In summary, we have derived a set of model equations that are general enough to be applicable to several driven systems with strain and magnetization as the order parameters. For instance, they capture a host of experimental results on the dynamics of driven magnetostrictive  metallic glass ribbon \cite{Vohra91a, Vohra91b, Vohra93, Vohra94, Vohra95}. Two other physical systems considered are the dynamics of driven magnetoelastic beam and martensitic ribbon. In all the  cases considered, the samples are thin long ribbons fixed at one end and driven at the other end.     Since in each case, the geometry of the experimental set-up is essentially the same, a natural starting point is to derive a general set of equations that can be reduced to each of these cases. This is what has been achieved here.  We have demonstrated that these set of equations do in fact recover a large number of experimental results.

In their most  general form, these equations  explain several experimental results on the strain dynamics of the magnetostrictive metallic glass ribbons  reported almost two decades ago \cite{Vohra91a, Vohra91b, Vohra93, Vohra94, Vohra95}. Detailed numerical studies of the model equations show that these equations reproduce the  period doubling route to chaos as a function of $h_{dc}$ keeping $h_{ac}$ fixed and also the quasiperiodic route to chaos as a function of $h_{ac}$ for a fixed dc field. The model equations predict induced and suppressed chaos when the system is subjected to an additional small amplitude near resonant perturbation with a frequency  $\Omega_r$ close to the first  subharmonic.  In our case, the suppressed chaos is seen for small $\delta$ ($\delta  = \Omega/2 - \Omega_r$) while induced chaos is seen for large $\delta$, opposite of what is reported in experiments.  This may be attributed to parametric type of forcing in our equations. These equations also exhibit symmetry restoring crisis with an exponent close to unity. 
We have also examined the influence of  near resonant forcing on the period doubling sequence as a function $h_{dc}$. We find that the entire bifurcation diagram is affected even for small $\delta$ consistent with the earlier report \cite{Chiz97}. While the successive PD bifurcations  leading to chaos are hastened, the nature of bifurcations for larger values of $h_{dc}$ are significantly altered in detail. However, for large $\delta$, the effect of resonant forcing is to induce modulation of the  amplitude of the original signal. For example the periodic orbits are modulated with a period $2\pi/\delta$ inducing large dispersion in  the Poincar\'e maps. 

The model equations have been adopted to explain the phase plot and the associated Poincar\'e map of a magnetoelastic beam driven at a low frequency \cite{Cusumano95}. We note here that these authors represent the coupling between magnetization and strain variables by providing a parametric drive for the strain variable, instead of describing the magnetic degrees of freedom along with the elastic degrees of freedom as has been done here. Clearly,  our set of equations are better suited to explain the experimental results as our model includes both strain and  magnetization. 

The model when reduced to describe the driven martensitic ribbon also  explains  the multiperiodic nonlinear response reported  in the internal friction experiments  on samples of nonmagnetic martensite samples of $Cu_{82.9}Al_{14.1}Ni_3$ \cite{Suzuki80, Wuttig81}. It is pertinent to note here that our equations automatically describe the twinned structure  in one dimension when the magnetic degree of freedom is dropped \cite{Bales}.  More interestingly, the equation for the strain amplitude exhibits some unusual switching dynamics between two equivalent wells of the periodic one orbit, which eventually makes a direct transition from period one cycle to chaos.  

Clearly, these equations can be easily adopted to describe the dynamics of driven magnetomartensites  as well \cite{FM}.   Indeed, rich dynamics is predicted in magnetomartensites where both elastic and magnetic nonlinearities are significant if experiments are performed in a similar geometry. While no such experiment has been reported so far, it would be interesting to verify this prediction.  We hope that the present work will encourage such dynamical studies. 

The approach is clearly applicable to any class of materials with two order parameters \cite{Fiebig02}.  For example these equations should describe the dynamics of driven ferroelectric materials that can be described by strain and electric polarization with the latter replacing  magnetization. Such systems are also described by polynomial forms of free energies as used in the model.  Our analysis suggests rich dynamics if experiments are performed in a similar geometry on material with two order parameters. 

Here it is worthwhile to point out that Eq. (\ref{S}) can be adopted to describe an order parameter where inertia is unimportant by taking $\gamma$ large. Then, it is clear that our approach can be easily extended to the case of ferroelectromagentic materials (such as  $RMnO_3$ where $R= Y, In$) \cite{Fiebig02}.   We hope that our work would encourage dynamic experiments in such  materials  where experiments are traditionally carried out in quasistatic conditions.

\begin{acknowledgments}
 {Acknowledgment: GA acknowledges the support from Indian National Science Academy for the Senior Scientist position. This work was supported BRNS Grant No: 2007/36/62-BRNS/2564.}
\end{acknowledgments}

\end{document}